\documentclass[twocolumn]{aastex631}
\usepackage{lineno}

\shorttitle{Resolved Gaia Triples}

\begin{document}

\renewcommand{\topfraction}{1.0}
\renewcommand{\bottomfraction}{1.0}
\renewcommand{\textfraction}{0.0}

\newcommand{\kms}{km~s$^{-1}$\,}
\newcommand{\msun}{$M_\odot$\,}
\newcommand{\masyr}{mas~yr$^{-1}$}

\title{Resolved Gaia Triples}

\author{Andrei Tokovinin}
\affiliation{Cerro Tololo Inter-American Observatory | NSF's NOIRLab
Casilla 603, La Serena, Chile}
\email{andrei.tokovinin@noirlab.edu}

\begin{abstract}
A sample  of 392 low-mass  hierarchical triple stellar  systems within
100 pc resolved  by Gaia as distinct sources is  defined. Owing to the
uniform  selection, the  sample is  ideally suited  to study  unbiased
statistics of wide triples.  The median projected separations in their
inner and  outer pairs are 151  and 2569 au, respectively,  the median
separation   ratio  is   close  to   15.   Some   triples  appear   in
non-hierarchical configurations, and many are just above the dynamical
stability limit.   Internal motions  in these  systems are  known with
sufficient accuracy to determine the orbital motion sense of the outer
and  inner pairs  and to  reconstruct the  eccentricity distributions.
The  mean  inner  and   outer  eccentricities  are  0.66$\pm$0.02  and
0.54$\pm$0.02,  respectively;  the  less eccentric  outer  orbits  are
explained by dynamical  stability.  The motion sense of  the inner and
outer pairs is almost uncorrelated, implying a mean mutual inclination
of  $83\fdg1  \pm  4\fdg5$.   The  median mass  of  the  most  massive
component is 0.71  \msun, the median system mass is  1.53 \msun.  In a
0.69 fraction of  the sample the primary belongs to  the inner binary,
while in the remaining systems it is the tertiary.  A 0.21 fraction of
the inner subsystems  are twins with mass ratios  $>$0.95.  The median
outer mass  ratio is 0.41;  it decreases mildly with  increasing outer
separation.   Presumably,  these  wide   hierarchies  were  formed  by
collapse   and  fragmentation   of  isolated   cores  in   low-density
environments and  represent a small  fraction of initial  systems that
avoided dynamical  decay.  Wide pre-main sequence  multiples in Taurus
could be their progenitors.
\end{abstract}

 \keywords{binaries:general --- binaries:visual --- star:formation}


\section{Introduction}
\label{sec:intro}

Binary stars and high-order  hierarchical systems are typical products
of  star  formation.  Parameters  of  stellar  systems (periods,  mass
ratios, eccentricies)  provide valuable insights into  star formation,
environment, and early  evolution, and serve as  benchmarks to compare
with  simulations  \citep[e.g.][]{Bate2014,Lomax2015}. This  fact  has
been  widely  recognized  regarding  binary  stars,  motivating  their
numerous      statistical     studies      \citep[][and     references
  therein]{DK13,Moe2017}. Stellar hierarchies containing three or more
stars are even more interesting from  this point of view, but they are
studied  much less.   The  reason is  a lack  of  unbiased samples  of
hierarchies because their discovery  is difficult and usually requires
combination of various techniques  such as high-resolution imaging and
spectroscopy.   A   relatively  complete  census  of   hierarchies  is
available for  solar-type stars  within 25\,pc \citep{R10},  but their
total  number  is  only  56.   Extension  of  this  effort  to  67\,pc
\citep{FG67a} or to smaller  masses \citep{Winters2019} is hampered by
the lack  of a  uniform coverage  of the parameter  space in  the full
range of periods and mass ratios.

The unprecedented census of stars provided by Gaia \citep{Gaia1,Gaia3}
offers an opportunity to gain  new information on stellar hierarchies,
extending recent works on wide Gaia binaries \citep[e.g.][]{EB2021}. I
use  here the  Gaia  Catalog  of Nearby  Stars  (GCNS) within  100\,pc
\citep{GCNS}  and select  from  it resolved  triple  stars.  The  Gaia
detection  capability  in terms  of  angular  separation and  contrast
restricts  this  sample  to  relatively wide  hierarchies  with  inner
separations on the  order of 100\,au.  On the other  hand, the GCNS is
complete  down  to the  lowest  stellar  masses, so  this  objectively
selected  sample  of Gaia  triples  gives  an  unbiased view  of  wide
low-mass  hierarchies.   This  is  a decisive  advantage  compared  to
samples of hierarchies  compiled from the literature  that suffer from
poorly known selection biases.  A collection of diverse hierarchies in
the Multiple Star Catalog, MSC \citep{MSC},  has been used in the past
for lack of better alternatives.

Most inner  subsystems in triples and  higher-order hierarchies within
100 pc are not resolved by  Gaia. Their existence can be inferred from
the excessive astrometric  noise, variable radial velocity  (RV), or a
flux  variability   caused  by  eclipses.   Moreover,   binaries  with
separations between 0\farcs1 and 0\farcs5 and comparable components do
not have Gaia parallaxes and hence  are missed in the GCNS, precluding
the study  of wide triples  containing such subsystems.  Only  a small
fraction of  all triples  are wide  enough to be  detected by  Gaia as
three distinct sources: they constitute 0.1 per cent of the total GCNS
population.  I  study here  this subset of  resolved wide  triples and
assume that each system contains  only three stars. This sample should
be relatively  complete above  the Gaia  separation-contrast detection
limit.

So, what can be learned from a complete sample of wide triples?  Their
masses and mass ratios inform us on the pairing probability and extend
to  triples the  known fact  that masses  of stellar  systems are  not
chosen  randomly even  at very  wide separations  \citep{EB2019b}. The
separation  ratios and  the sense  of orbital  motion can  distinguish
products  of  core fragmentation  from  products  of chaotic  dynamics
(e.g. decay of  small clusters). Furthermore, the  direction and speed
of relative  motion constrain  the eccentricity  distributions.  Taken
together,  these  observational facts  help  us  to better  understand
formation of hierarchical systems.

The  sample  of   wide  triples  is  defined   and  characterized  in
section~\ref{sec:sample}.   Its  statistical   properties  (separation
ratios, relative  sense of motion, eccentricities,  and mass ratios)
are  presented in  section~\ref{sec:stat}.  The origin  of these  wide
triples is  discussed in section~\ref{sec:disc}, and  the main results
are summarized in section~\ref{sec:sum}.

\section{The Sample}
\label{sec:sample}

\subsection{Relative Velocities in Wide Triples}
\label{sec:vel}

\begin{figure}[ht]
\centerline{
\includegraphics[width=8.5cm]{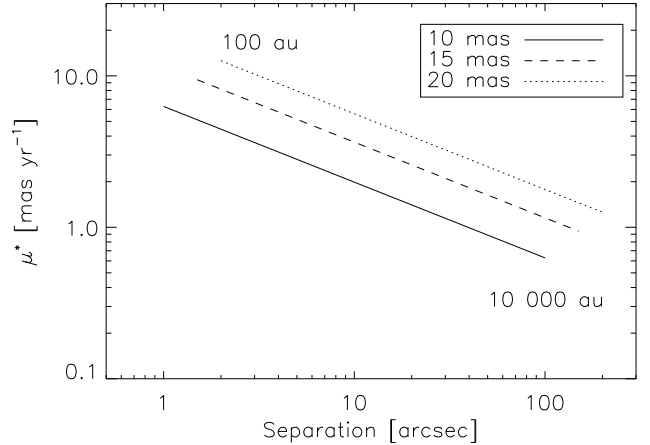} }
\caption{Characteristic   orbital  speed   of  binaries   vs.  angular
  separation  for three  values  of parallax  and the  mass  sum of  1
  \msun. The curves span projected separations from 100 to $10^4$ au.
\label{fig:diagplot} 
}
\end{figure}

Wide outer  subsystems with  separations of $10^3$  to $10^4$  au move
slowly, but  the accuracy of  the Gaia  proper motions (PMs)  is still
sufficient to detect these motions out to a distance of $\sim$100\,pc.
The orbital speed (in arcsec~yr$^{-1}$)  of a binary with a separation
$\rho$ (in arcseconds) and a face-on  circular orbit with period $P$
is
\begin{equation}
\mu^* =  (2 \pi \rho)/P = 2 \pi \rho^{-1/2} \varpi^{3/2} M^{1/2},
\label{eq:mu*}
\end{equation}
where $\varpi$ is  the parallax in arcseconds and $M$  is the mass sum
in  solar units.  This parameter,  called characteristic  speed, is  a
scaling  factor  for  binaries   with  arbitrary  eccentricity,  orbit
orientation, and  phase. For  a bound binary,  the relative  speed is
always less  than $\sqrt{2} \mu^*$, and a typical speed
is about  half of  $\mu^*$  \citep{Tok2016}. In the  following, I  use the
scaled relative speed $\mu' = \Delta \mu / \mu^*$. 

Figure~\ref{fig:diagplot} plots  eq.~\ref{eq:mu*} for three  values of
parallax. Most PM errors in our sample are smaller than 0.1 \masyr, so
even for the widest and most  distant pairs the relative motion can be
measured  with  a   signal  to  noise  ratio  above  3,   with  a  few
exceptions. However,  further extension  of the maximum  separation or
distance  will  be  restricted  by the  accuracy  and  reliability  of
relative motions deduced from Gaia.

\subsection{Selection Criteria}
\label{sec:crit}

I searched the full GCNS catalog of 331312 stars to identify groups of
three or more stars that are located close in space and share a common
PM. The selection criteria are:
\begin{itemize}
\item
Parallaxes equal within 1\,mas.
\item
Projected separation $s < 10^4$ au.

\item
Relative projected  speed (in  \kms) $\Delta  V <  10 (10^3/s)^{0.5}$,
where $s$ is expressed in au.  This is a relaxed form of the boundness
criterion  which   rejects  optical   companions  but   preserves  the
hierarchies. The strict criterion $\mu' < \sqrt{2}$ is applied later.

\item
Masses less than 1.5 \msun to avoid evolved stars.

\end{itemize}

The search returns 840  resolved hierarchies within 100\,pc, including
30 quadruples  and one quintuple,  $\xi$~Sco.  The cumulative  plot of
the  number  of  hierarchies  vs.   distance limit  $d$  can  be  well
approximated by  the $d^{2.2}$  law.  A  cubic law  is expected  for a
complete  sample.   However, with  the  increasing  distance the  Gaia
resolution  limit  removes  a  progressively larger  number  of  inner
subsystems  and slows  the growth.   The number  of hierarchies  found
within 67\,pc is 342, and only  156 of those were previously known and
listed in the MSC. The 186 new hierarchies within 67\,pc were added to
the MSC.   I have not yet  added the remaining new  hierarchies out to
100\,pc; most of  them are composed of faint  stars without additional
information in the literature.

\begin{figure}[ht]
\includegraphics[width=8.5cm]{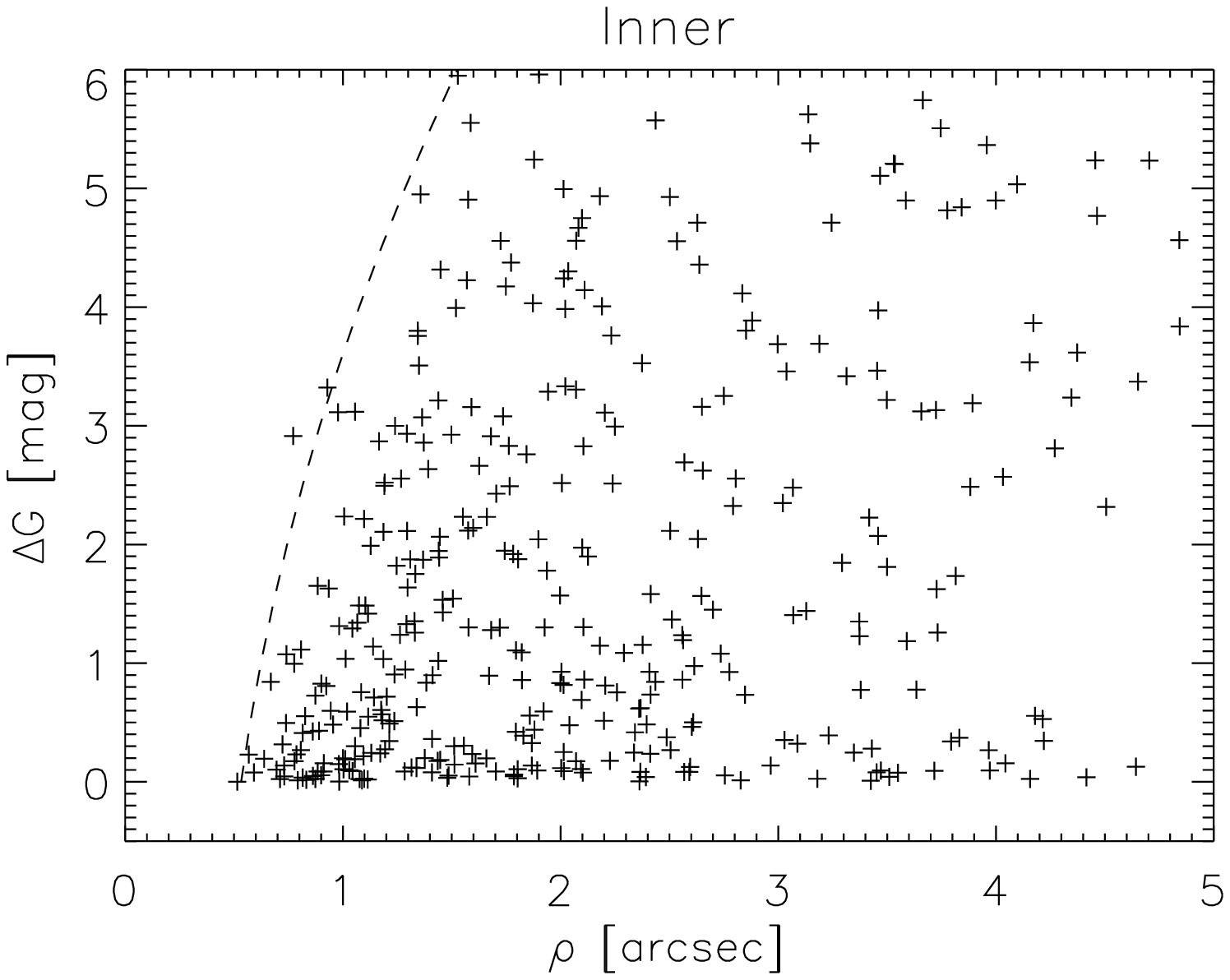} 
\includegraphics[width=8.5cm]{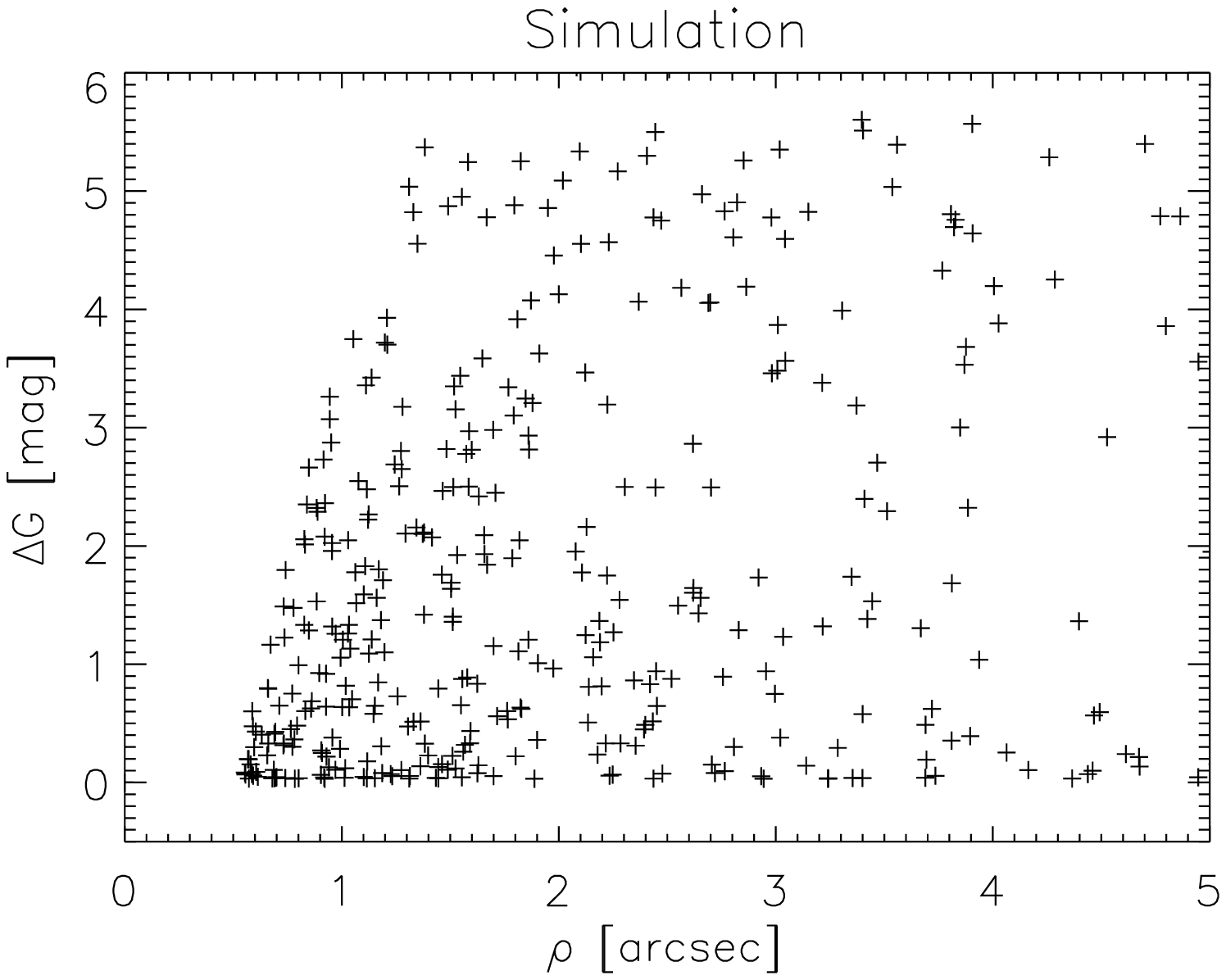} 
\caption{Magnitude difference $\Delta G$ vs. separation $\rho$ for
 the inner subsystems in the actual (top) and simulated (bottom)
 samples. The dashed line shows the nominal Gaia detection limit.
\label{fig:sep-dg} 
}
\end{figure}

According to the eq.~2 in \citet{GCNS}, the detection of companions is
complete at an angular separation of  0\farcs53 for equal pairs and at
1\farcs28 for  pairs with $\Delta  G < 5$ mag,  or a mass  ratio above
$\sim$0.4.   The top  plot in  Figure~\ref{fig:sep-dg} shows  that the
Gaia detection curve (dashed line)  matches well the upper envelope of
the inner  subsystems. However, when  a realistic population  of inner
binaries is simulated (see  section~\ref{sec:fe}) and filtered by this
limit, the number of points near the  curve is larger than in the real
sample, indicating that Gaia misses some close pairs within the stated
limit.

\subsection{Simple Triples and the Impact of Subsystems}
\label{sec:subsystems}

The members of selected hierarchies  are not necessarily simple stars,
as each of their components  can contain a close unresolved subsystem.
As  noted,  GCNS  does  not  include  resolved  close  binaries,  thus
eliminating some subsystems. I removed  from the sample all stars with
RUWE  (reduced  unit  weight  error)   above  2  that  likely  contain
subsystems with periods from a  few years to several decades, revealed
by  deviations   from  the  5-parameter  astrometric   model  used  in
Gaia.  Note that  some  inner subsystems  have  an increased RUWE  simply
because of their non-linear orbital motion.

Inner subsystems with  periods of $\sim$100\,yr or longer  can cause a
quasi-linear displacement of the photocenter during the 3-yr Gaia time
window  that is  not  detectable  by the  increased  RUWE. Assuming  a
subsystem  with  a  semimajor  axis  of 20  au  at  100\,pc  distance,
eq.~\ref{eq:mu*} gives $\mu^* = 14$ \masyr. A typical orbital speed is
two  times less  than  $\mu^*$, and  the  photocenter amplitude  gives
another factor of 3 reduction compared  to the relative speed, but the
remaining bias of $\sim$2 \masyr  ~is still substantial and comparable
to    the    expected    motion    in   wide    outer    pairs    (see
Figure~\ref{fig:diagplot}).  As  a result,  the outer pair  may appear
unbound.  I  removed from  the sample  48 hierarchies  with apparently
unbound (mostly outer) pairs with $\mu' > \sqrt{2}$, likely containing
subsystems.  Also  removed are  15 hierarchies where  inner subsystems
are documented in  the MSC, as well as the  hierarchies with more than
three stars in the GCNS.  The remaining 392 hierarchies are considered
here  as simple  triples without  inner subsystems  and represent  the
sample used for the statistical studies below.

Undetected  inner subsystems  certainly remain  in  some of  the
hierarchies in this  sample. Subsystems with periods  below $\sim$1 yr
do not perturb the relative motion  (their effect is averaged over the
Gaia   time-span),  they   only  increase   the  mass   sum  $M$   and,
correspondingly, $\mu^*$.   Most subsystems with  intermediate periods
are  filtered  out  by  the   RUWE  criterion,  and  only  long-period
subsystems can distort the measured relative motion, in the worst case
by as much as a few \masyr.  Their effect is to increase $\mu'$ and to
add random errors to the direction  of relative motion, similar to the
effect  of the  measurement  errors.  I  neglect potential  undetected
subsystems in the following.

\subsection{Distribution of Masses}
\label{sec:masses}

\begin{figure}[ht]
\includegraphics[width=8.5cm]{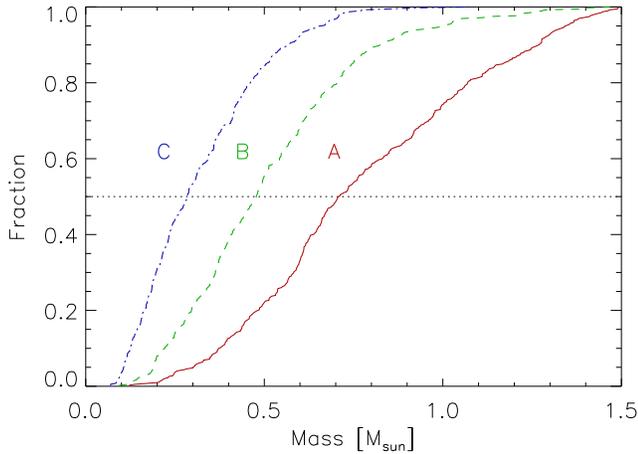} 
\caption{Cumulative distribution of masses in the components of the
  triples ordered from the brightest (A) to the faintest (C).
\label{fig:mass} 
}
\end{figure}

Masses  of  the individual  components  of resolved  triples  are
estimated from  the absolute magnitudes  $M_G$ in the Gaia  band using
the  1-Gyr solar-metallicity  PARSEC isochrone  \citep{PARSEC}.  Stars
brighter than $M_G = 2.66$ mag (some of which are evolved) are removed
in the  process, so all  triples have  components on  the main
sequence with masses  less than 1.5 \msun. The lowest  stellar mass of
0.08  \msun corresponds  to  $M_G =  16.7$  or $G=  21.7$  mag at  a
distance of  100\,pc (distance modulus 5  mag).  This is close  to the
Gaia magnitude limit,  so the GCNS samples all stellar  masses down to
the hydrogen-burning limit.  Only 14 triples (3.6 per cent) have their
smallest  component less  massive  than 0.1  \msun.   There are  known
triple  brown dwarfs  in the  field (e.g.   DENIS J020529.0-115925  at
20\,pc,  separations 0\farcs4  and  0\farcs07)  and brown-dwarf  twins
orbiting more massive  stars (e.g.  the 0\farcs13  brown-dwarf pair at
90\arcsec ~from HD~97334).  However, these  triples are not resolved by
Gaia and are not present in our sample.

The median  absolute magnitude of the  GCNS stars $M_G =  10.8$ (which
coincides with the  peak of the luminosity function)  corresponds to a
star of 0.32 \msun. The median mass of all stars belonging to our
wide triples is  0.47 \msun, so multiples are on  average more massive
than single  stars.  The  minimum, median, and  maximum values  of the
total (system) mass are 0.35, 1.53, and 3.70 \msun, respectively.

Figure~\ref{fig:mass}  shows  the  cumulative distribution  of  masses
labeled as A,  B, C in order of decreasing  brightness in each triple.
The median values are 0.71, 0.49, and 0.29 \msun, respectively, so the
resolved triples  studied here  are composed mostly  of K-  and M-type
dwarfs.  For  comparison, I  simulated triples selected  randomly from
the  mass function  of  the GCNS  sample.  The  median  masses in  the
simulated sample are smaller, 0.59,  0.31, and 0.16 \msun.  The ratios
of median tertiary  to median primary masses are 0.27  and 0.41 in the
simulated  and real  samples,  respectively.  This  result extends  to
triples  the  property of  wide  binaries  to have  non-random  masses
\citep{EB2019b}.

\begin{figure}[ht]
\includegraphics[width=8.5cm]{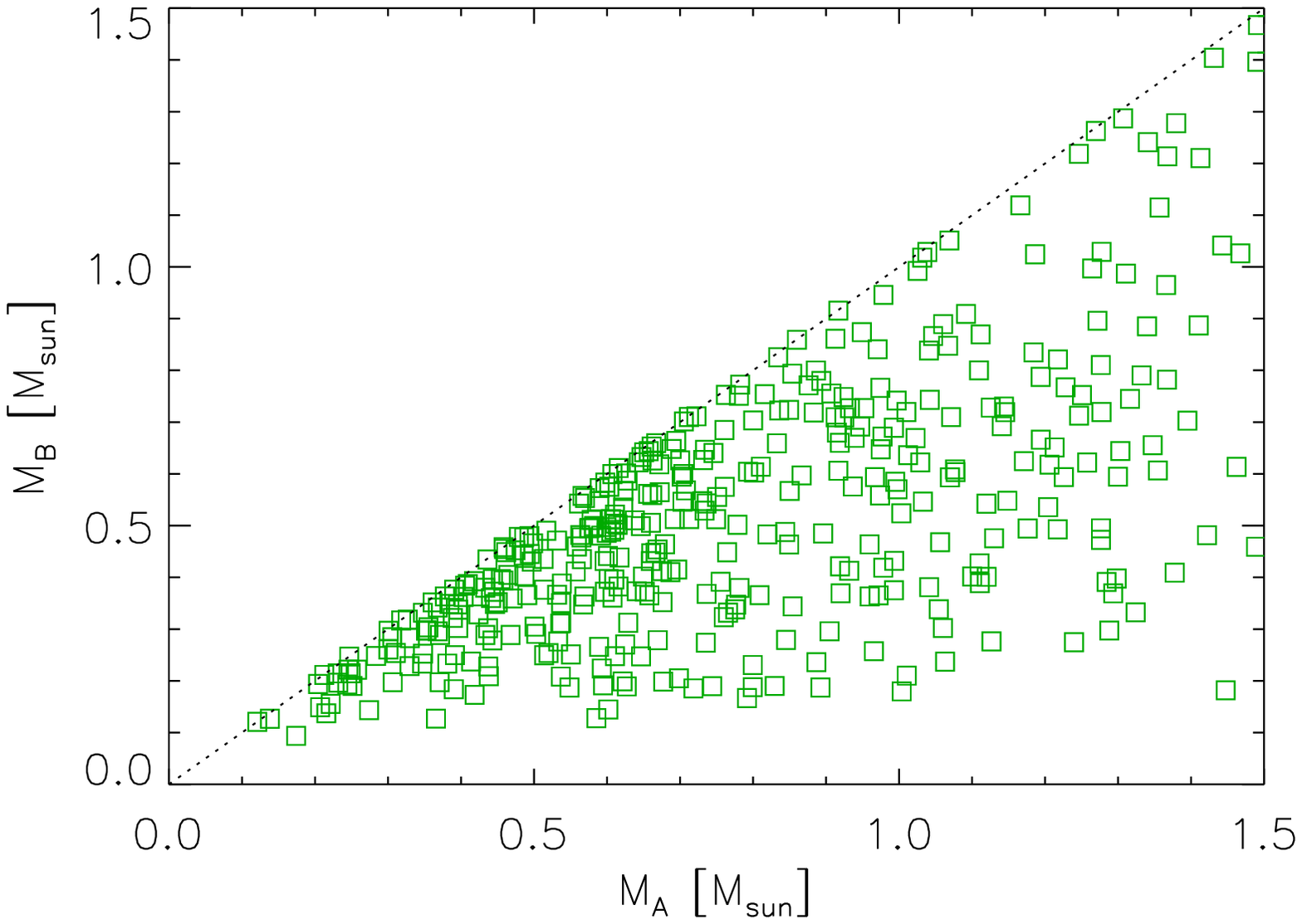} 
\includegraphics[width=8.5cm]{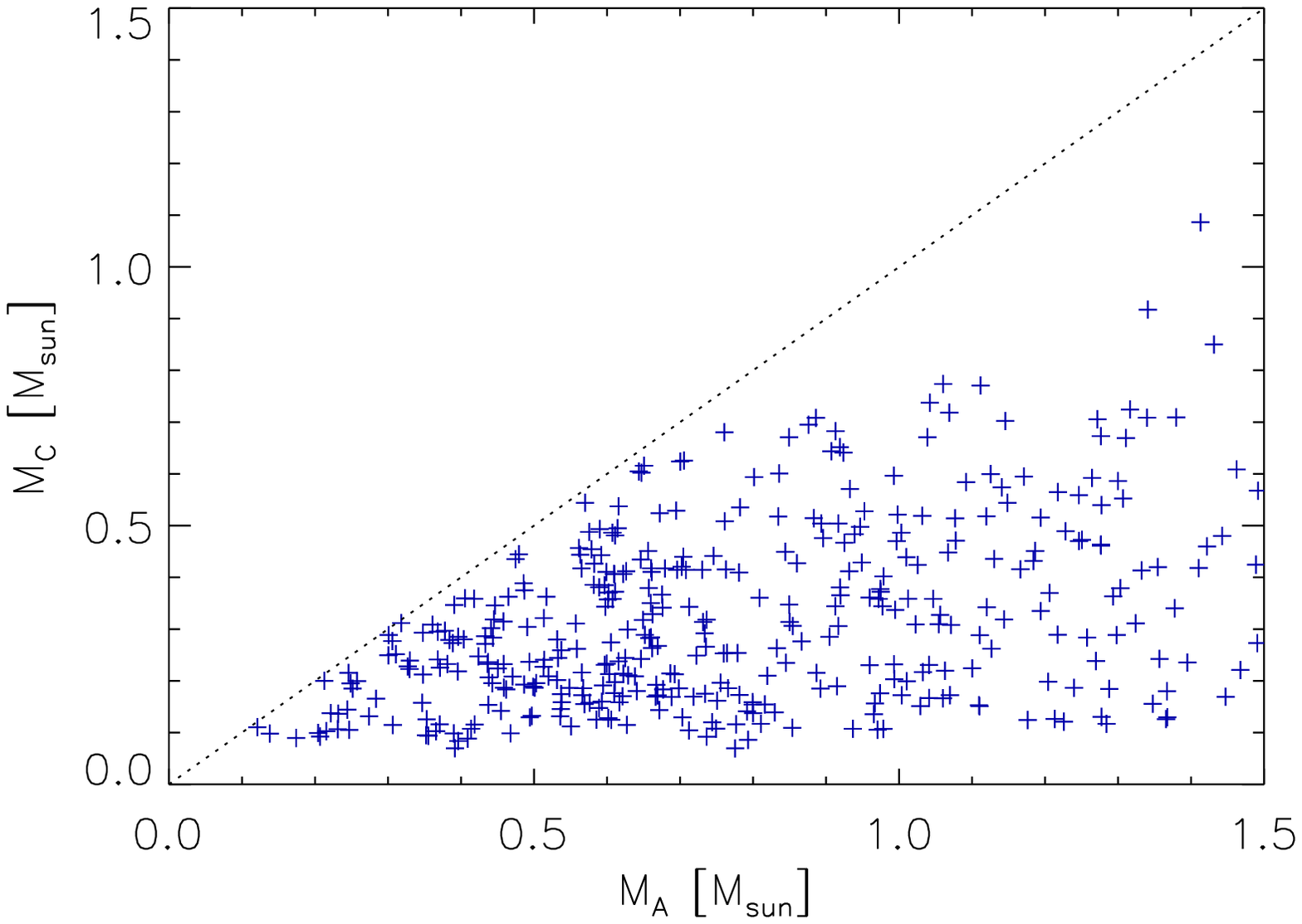} 
\caption{Relation between masses of the brightest star A and the
  second brightest star B (top) or the dimmest star C (bottom). The
  dotted line corresponds to mass equality.
\label{fig:massplot} 
}
\end{figure}

The plots in Figure~\ref{fig:massplot} compare masses of the components
B  and C  with the  mass of  the largest  component A.   The two  most
massive components A  and B often are similar (twins),  as revealed by
grouping of the points near  the diagonal.  The majority of equal-mass
pairs  belong to  the inner  subsystems.  However,  triples where  all
three stars have similar masses, $M_C  \approx M_A$, are found only at
$M_A  < 0.75  $  \msun.   The triples  studied  here  are wide  (inner
separations $\sim$100  au).  Close (e.g.  spectroscopic)  triples with
three nearly-equal  solar-mass components are quite  common, and their
absence at  large separations  tells us  something about  formation of
hierarchical systems. 

\subsection{Kinematics of Resolved Triples}
\label{sec:kin}

\begin{figure}[ht]
\includegraphics[width=8.5cm]{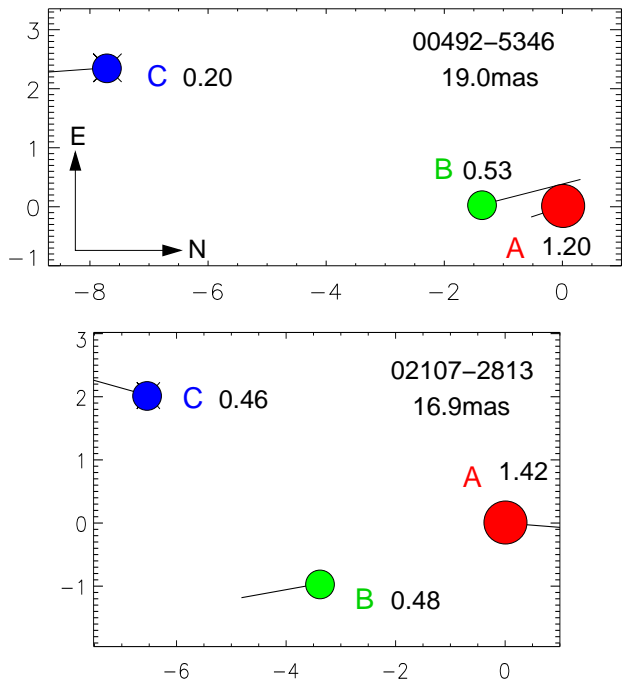} 
\caption{Two resolved  triples on  the sky,  scale in  arcseconds. The
  red, green,  and blue  circles denote stars  in order  of decreasing
  mass, the  masses are indicated  by numbers. Lines coming  from each
  star show its relative PM, reaching $\sim$5 \masyr in both
  cases. The WDS codes and mean parallaxes are indicated. 
\label{fig:sky} 
}
\end{figure}

Placing the  most massive star A  at the coordinate origin,  I compute
the rectangular  X,Y coordinates of  other stars (X directed  North, Y
East)  and   the  differential   PMs  after  subtracting   the  common
center-of-mass PM.  Figure~\ref{fig:sky}  illustrates two triples. The
first  is quite  typical, the  second has  comparable separations  and
appears as non-hierarchical. Given the coordinates, pairwise distances
between the stars are computed.  The closest pair is considered as
inner  (independently  of  masses), and the remaining  star  is the tertiary
component.   This  sets the  inner  and  outer separations  $\rho$  in
arcseconds and the corresponding  position angles $\theta$. The angles
are counted counter-clockwise  from North. The median  inner and outer
angular separations are 151 and 2570 au, respectively.

The differential PM in the inner  pair serves to compute the magnitude
$\Delta  \mu$ and  the  position angle  $\theta_\mu$  of the  relative
motion vector (secondary relative to  primary).  The angle $\gamma$ of
the  relative motion  with respect  to  the pair's  position angle  is
$\gamma =  \theta_\mu -  \theta$. This  angle is  defined in  the full
$360^\circ$ range.   Values between 0 and  $180^\circ$ indicate direct
(counter-clockwise)  orbital motion,  small  $\gamma$  means a  radial
motion with increasing separation,  while $\gamma \approx 90^\circ$ or
$\gamma \approx 270^\circ$ imply  orthogonal motion.  Referring to the
top panel  of Fig.~\ref{fig:sky},  the inner  close pair  has $\theta=
178.5^\circ$ and $\gamma = 196.8^\circ$ (stars approach each other and
move clockwise).  Using the center of mass of the inner pair, the same
parameters, relative position and $\gamma$, are computed for the outer
subsystem.   In this  case, $\gamma  = 21.9^\circ$,  meaning that  the
third star moves away and the outer pair moves counter-clockwise.  The
median errors of $\gamma$ are 1\fdg2 and 2\fdg9 in the inner and outer
subsystems,  respectively.   Only  10  triples have  errors  of  outer
$\gamma$  exceeding  $30^\circ$.   They   are  not  removed  from  the
statistical analysis to avoid potential bias.

\begin{deluxetable*}{r cc c lc cc  c cc cc  }
\tabletypesize{\scriptsize}     
\tablecaption{Data on resolved wide triple systems (fragment)
\label{tab:1} }  
\tablewidth{0pt}                                   
\tablehead{
\colhead{$N$} &                                                                     
\colhead{WDS} & 
\colhead{$\varpi$} &
\colhead{Comp.} &
\colhead{EDR3} & 
\colhead{$G$} & 
\colhead{$M$} & 
\colhead{$\rho$} & 
\colhead{$\theta$} & 
\colhead{$\gamma$} & 
\colhead{$\sigma_\gamma$} & 
\colhead{$\Delta \mu$} & 
\colhead{$\mu'$} \\
&
\colhead{(J2000)} & 
\colhead{(mas)} &
 & &
\colhead{(mag)} & 
\colhead{(\msun)} & 
\colhead{($''$)} & 
\colhead{(deg)} & 
\colhead{(deg)} & 
\colhead{(deg)} & 
\colhead{(\masyr)} & 
 }
\startdata
 12 & 00492-5346 &18.84 &A &4921910067405272320 & 7.46  &1.20   &      &      &      &      &      &  \\
 12 &            &19.19 &B &4921910067405761024 &12.41  &0.54   & 1.36 &178.5 &196.9 &  2.3 & 5.62 & 0.30 \\
 12 &            &18.83 &C &4921910063109280896 &15.83  &0.20   & 7.65 &162.2 & 21.9 &  4.3 & 3.55 & 0.43 \\  
 42 & 02107-2813 &16.94 &A &5117251502218279552 & 6.80  &1.42   &      &      &      &      &      &     \\
 42 &            &16.81 &C &5117251502218710784 &13.41  &0.46   & 3.50 &195.9 &347.6 &  0.6 & 6.96 & 0.69 \\
 42 &            &17.00 &B &5117251502217614848 &13.21  &0.48   & 6.12 &158.4 &  7.9 &  0.4 & 6.32 & 0.74 \\
\enddata
\end{deluxetable*}


\subsection{Description of the Catalog}
\label{sec:table}

Relevant  data  on the  392  wide  triple  systems  are given  in  the
electronic  Table~\ref{tab:1}. Its  fragment  reproduced  in the  text
refers  to  the  triples  in Figure~\ref{fig:sky}.  The  first  column
contains the internal  sequential number of the  system, replicated in
the two following lines (three  lines per system: inner primary, inner
secondary, and tertiary).  The WDS code  in column (2) is based on the
J2000 coordinates of  the brightest star, as in  the Washington Double
Star catalog \citep{WDS}.   Column (3) contains the  parallaxes of the
stars, column (4) gives the mass-related  component labels A, B, or C.
The Gaia EDR3 identifiers in column  (5) provide links to the Gaia and
GCNS catalogs,  so there  is no  need to  repeat the  full astrometric
information here.  For reference,  $G$ magnitudes and estimated masses
are given in columns (6)  and (7), respectively. The remaining columns
are empty  for the  inner primary.   For the  inner secondary  and the
tertiary, they contain separation  $\rho$ (8), position angle $\theta$
(9),  angle  $\gamma$ (10),  its  error  (11), relative  motion  speed
$\Delta \mu$ (12),  and its normalized equivalent $\mu'$  (13) for the
inner and outer  subsystems.  The errors of $\gamma$  are estimated by
the simplified formula $\sigma_\gamma  = 57\fdg3 \sigma_{\Delta \mu} /
\Delta  \mu$,  therefore  there  is  no need  to  provide  the  errors
$\sigma_{\Delta \mu}$.

\section{Statistics}
\label{sec:stat}

\subsection{Separation Ratio}

\begin{figure}[ht]
\includegraphics[width=8.5cm]{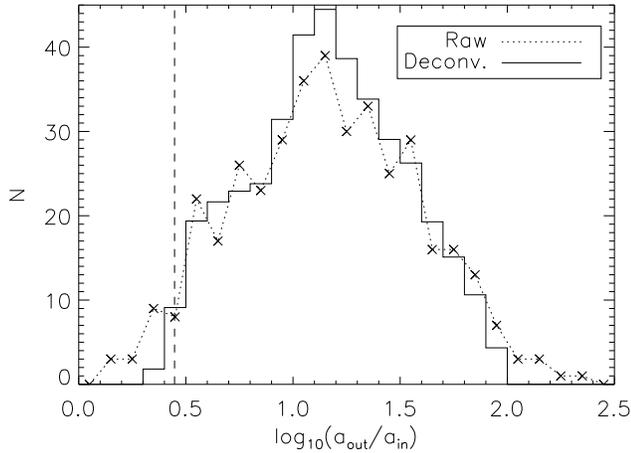} 
\caption{Distribution  of  the  separation   ratio  (dotted  line  and
  crosses) and  its deconvolution from  projections into the  ratio of
  semimajor axes (full histogram). The vertical line shows the minimum
  semimajor axis ratio of 2.8 allowed by dynamical stability. 
\label{fig:sepratio} 
}
\end{figure}

Several apparently non-hierarchical triples (see Figure~\ref{fig:sky})
raise      concern      about     their      dynamical      stability.
Figure~\ref{fig:sepratio}  shows the  distribution  of the  separation
ratio, often expressed as $x = \log_{10} s_{\rm out}/s_{\rm in}$.  The
median ratio  is 14.75, or $x  = 1.17$. Wide triples  cataloged in the
MSC \citep{MSC} have median $x=1.33$, and  some of them also appear as
non-hierarchical.  The  distribution of the projected  separations can
be deconvolved from projections following  the recipe described in the
Appendix of the MSC paper, converting  it into the distribution of the
semimajor  axis ratio.   As  expected,  it is  more  compact than  the
distribution  of the  separation ratio.   The right-hand  side of  the
distribution is shaped  by the selection (large ratios are  cut off by
the Gaia resolution  limit).  The left part of the  curve implies that
some   systems   may    be   below   the   minimum    ratio   of   2.8
\citep{Mardling2001}.  However, this could be caused by the smoothness
of the deconvolved distribution, imposed  to damp the noise.  There is
no firm evidence that truly unstable triples are present in the field;
such  systems decay  rapidly.  However,  existence of  many marginally
stable wide triples is unquestionable.

\subsection{Relative sense of orbital motion and mutual inclination}

\begin{figure}[ht]
\includegraphics[width=8.5cm]{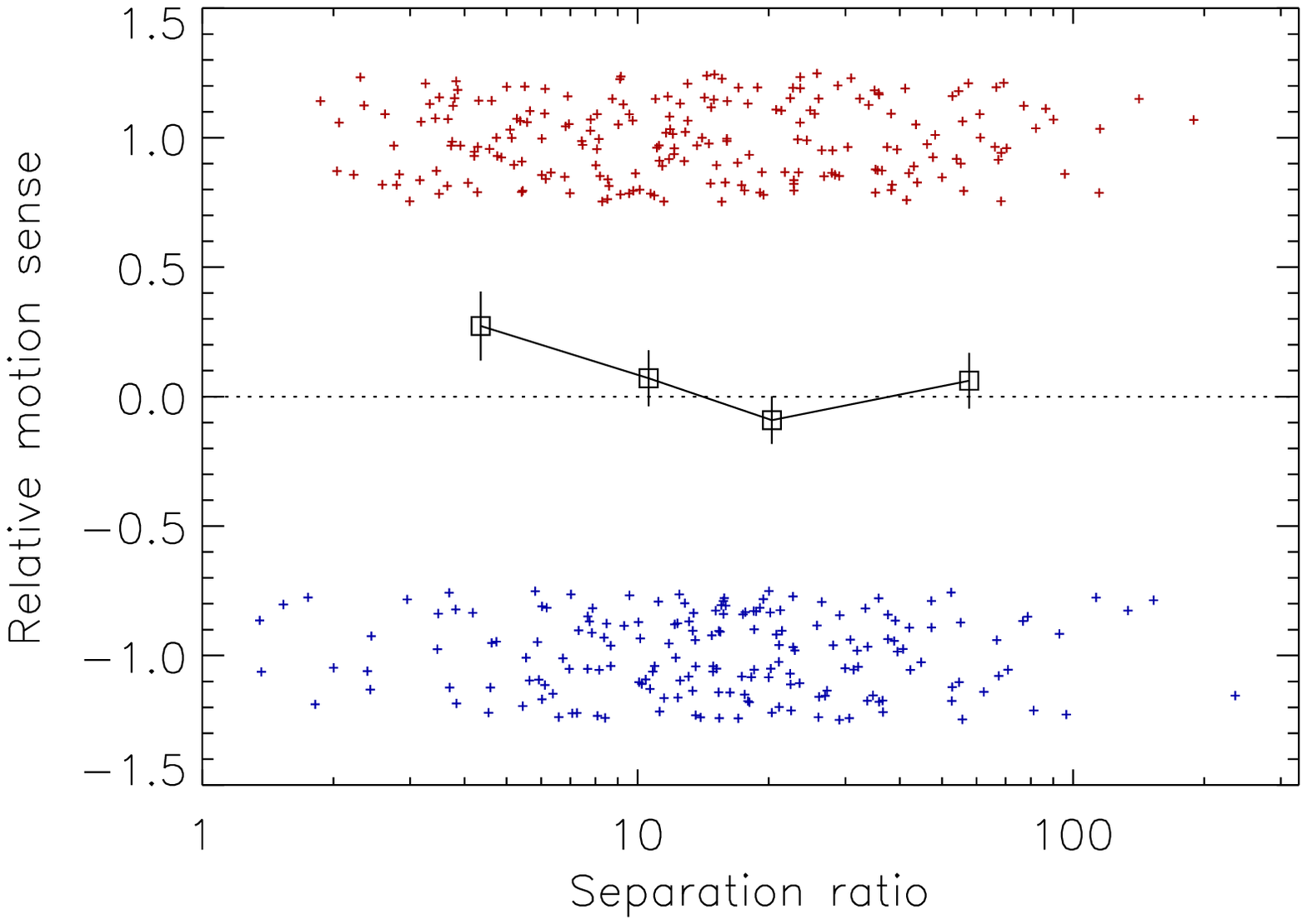} 
\includegraphics[width=8.5cm]{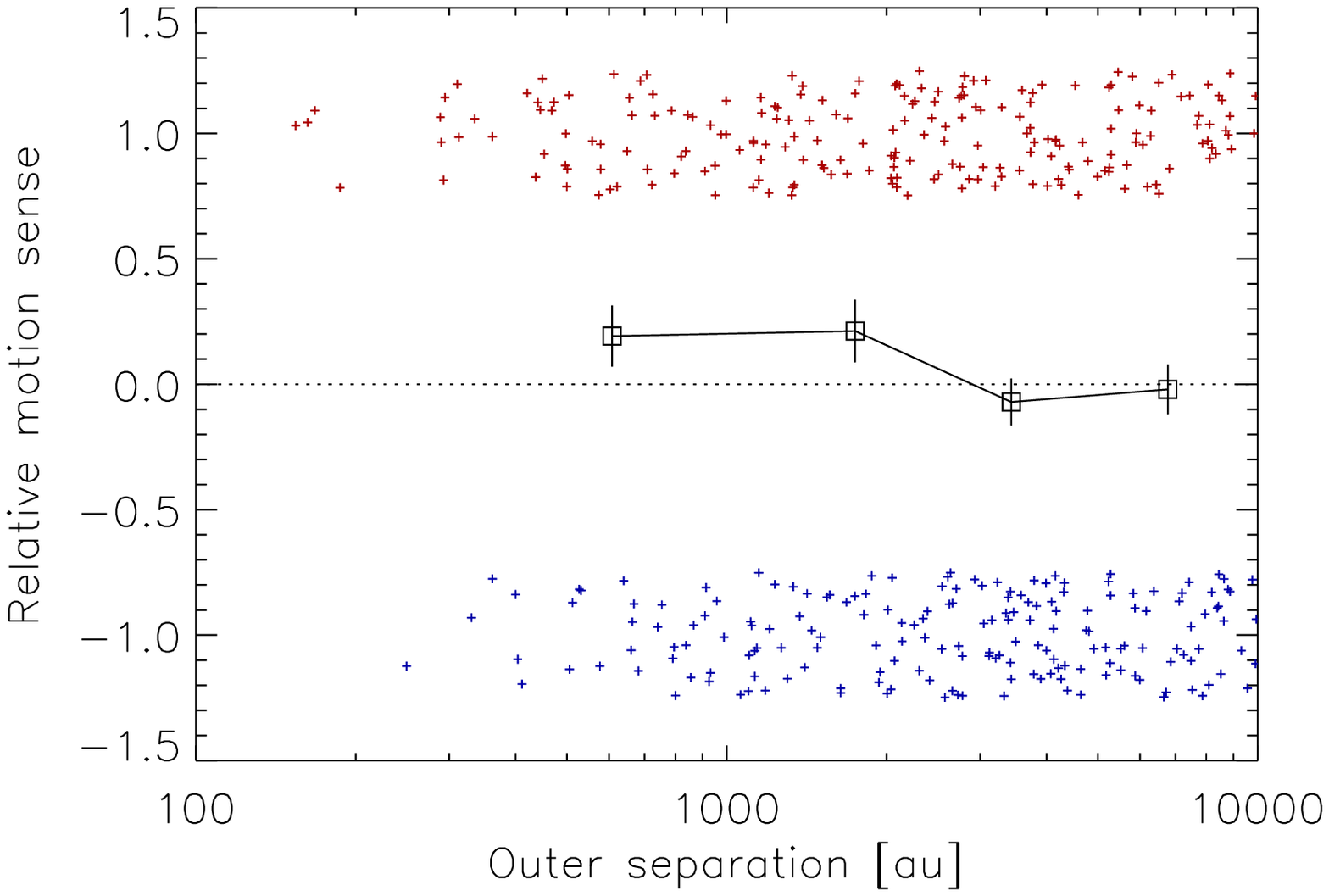} 
\caption{Relative sense  of orbital motion vs.  separation  ratio (top) and
  outer separation  (bottom).  The  values of +1 and $-$1 correspond
  to matching and opposite directions of motion, respectively. The
  symbols are randomly displaced vertically to avoid overlap. The
  squares with error bars are sign correlations $C$ computed for
  quarters of the sample. 
\label{fig:spin} 
}
\end{figure}

The inner and outer angles $\gamma$  define the sense of revolution in
these triples, allowing some inference  on the mutual orbit alignment.
Of the 392 triples,  $N_+ = 211$ revolve in the  same sense.  The sign
correlation $C = (N_+ - N_-)/(N_+  + N_-)$ is $0.076 \pm 0.050$.  This
parameter is related to the mean mutual inclination $\Phi = 90^\circ(1
- C) =  83\fdg1 \pm 4\fdg5$.  In  a sample of 216  visual triples from
the MSC,  a significant degree  of orbit  alignment has been  found at
outer separations  below 50 au, while  triples wider that 1000  au had
random  mutual  inclinations   \citep{Tok2017}.   To  probe  potential
dependence of mutual orbit alignment in wide triples on the separation
ratio  and  outer  separation,  I plot  in  Figure~\ref{fig:spin}  the
relative motion sense (positive and  negative numbers for triples with
orbital  motion in  the  same and  opposite  sense, respectively)  vs.
these parameters.  The sample was  split over each parameter into four
equal groups, and $C$ was computed  for each group (squares with error
bars).  One  can see  that more  compact triples seem  to have  a weak
(although not statistically significant) alignment, while wide triples
definitely have randomly oriented  orbits.  This confirms the tendency
found in  the MSC and extends  it to wider separations,  thanks to the
accurate Gaia astrometry.

\subsection{Eccentricities}
\label{sec:fe}

\begin{figure}[ht]
\includegraphics[width=8.5cm]{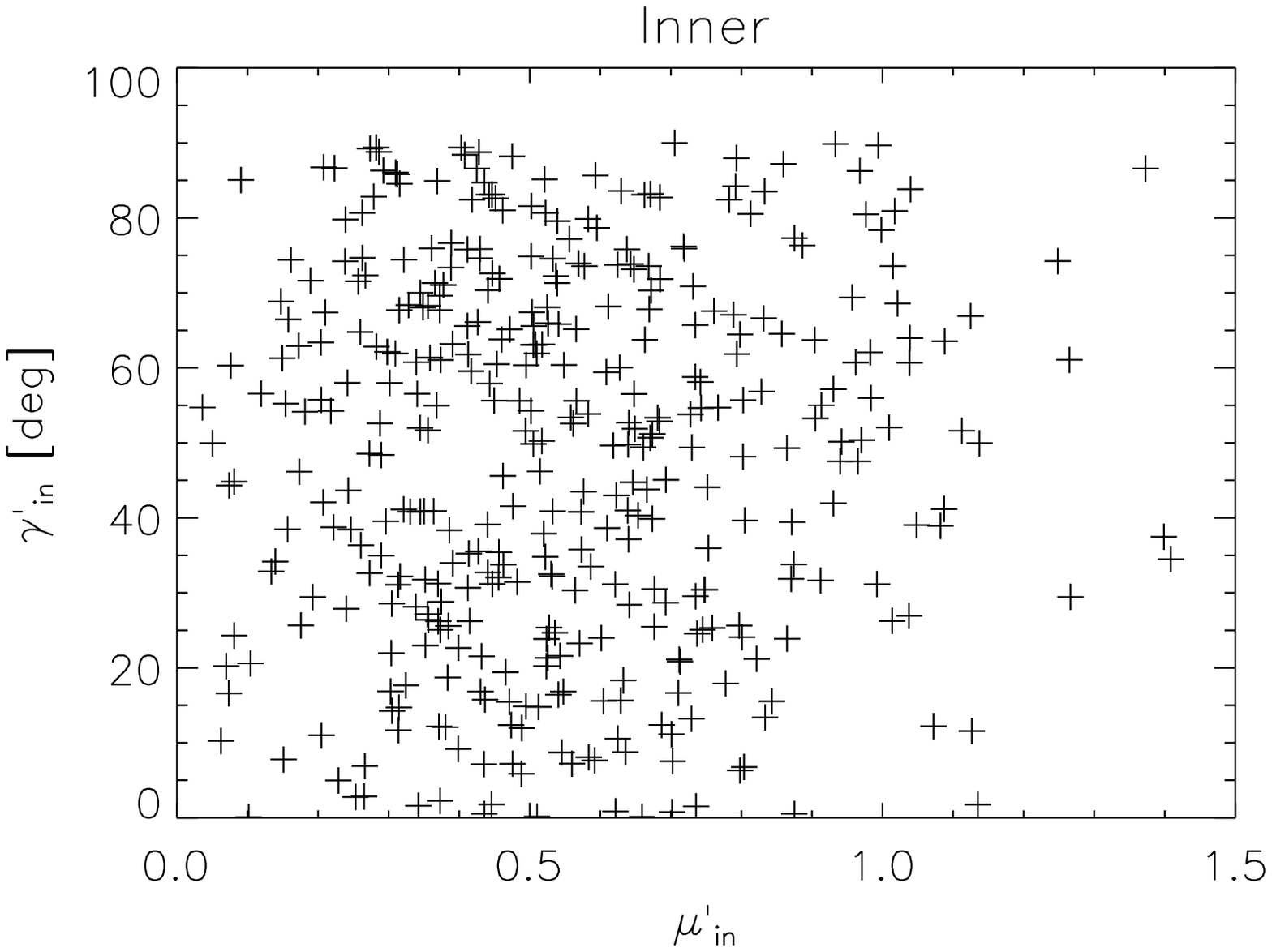} 
\includegraphics[width=8.5cm]{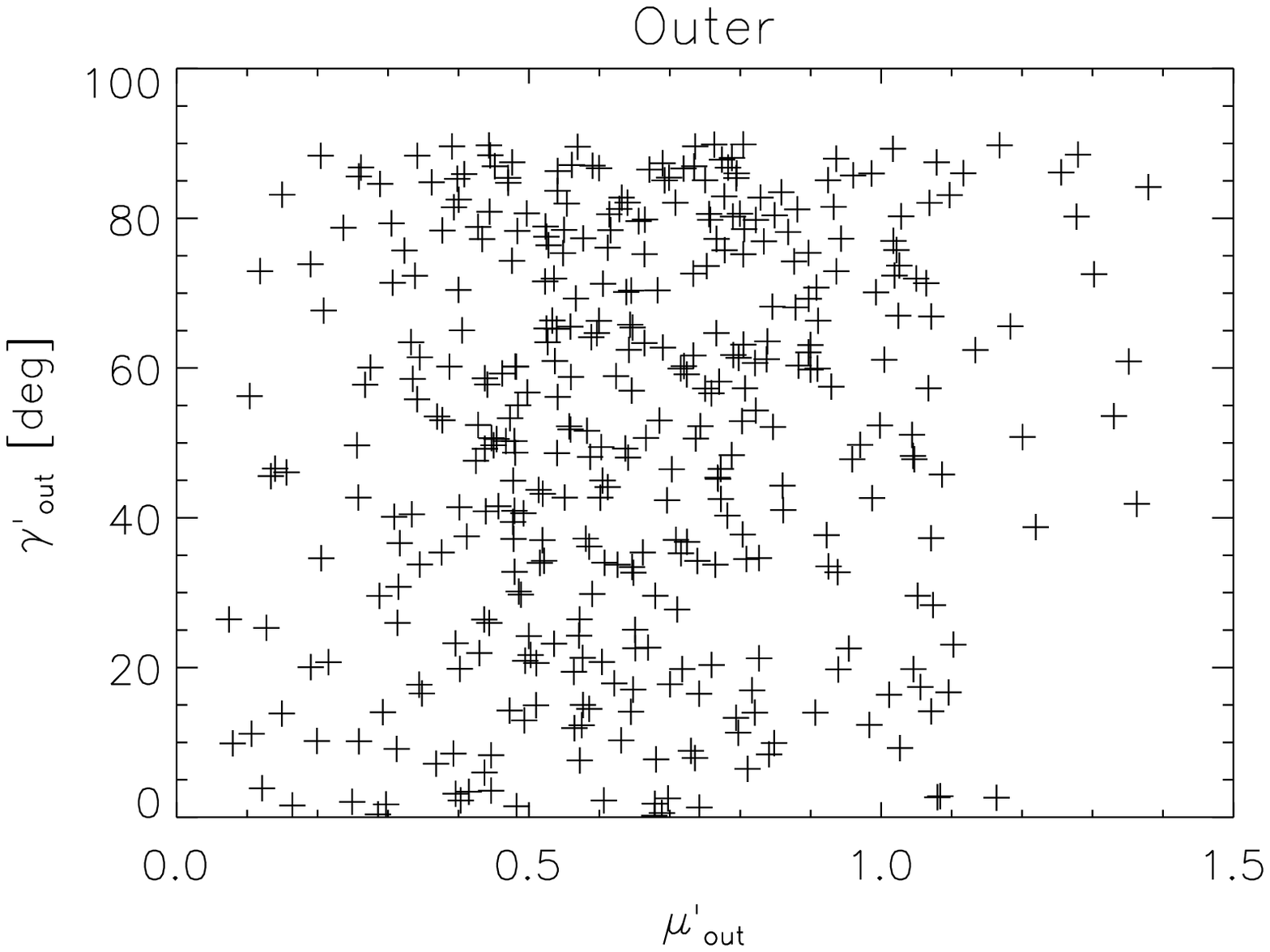} 
\caption{Plots of $\gamma'$ vs. $\mu'$ for the inner (top) and outer
  (bottom) subsystems. 
\label{fig:mu-gamma} 
}
\end{figure}

The joint distribution of $(\mu',  \gamma')$ allows to reconstruct the
eccentricity  distribution,  as  demonstrated by  \citet{Tok2016}  and
\citet{Tok2020}.  Here $\gamma'  = \arcsin ( |  \sin(\gamma)|)$ is the
angle     folded     into     the    $(0,     90^\circ)$     interval.
Figure~\ref{fig:mu-gamma} shows  these plots  for the inner  and outer
subsystems.  Although the  number of points is modest,  one notes that
values $\gamma' \sim  90^\circ$ are more frequent in  the outer pairs.
Similar plots simulated for  binaries with fixed eccentricities easily
distinguish     between     circular     and     eccentric     orbits.
Table~\ref{tab:gamma} gives  the median values of  the parameters (the
errors are determined by bootstrap) and their correlation coefficients
$\rho_{\mu \gamma}$.  Its  last two columns give these  values for the
thermal  and  uniform  eccentricity distributions,  as  determined  by
\citet{Tok2016}.   For   reference,  $e=0.8$  corresponds   to  median
$\gamma'  =  41\fdg4$  and  a  strong  anti-correlation  $\rho_{\mu
  \gamma} = -0.21$.

\begin{table}
\center
\caption{Statistics of relative motion}
\label{tab:gamma}
\medskip
\begin{tabular}{l   c  c c  c } 
\hline
Parameter &  Inner  & Outer  &  $f(e)=2e$ & $f(e)=1$ \\
\hline
Median $\gamma'$        & 50.4$\pm$2.2 &  53.3$\pm$2.6 & 45.0  & 53.5 \\ 
Median $\mu'$           & 0.52$\pm$0.04& 0.64$\pm$0.02 & 0.546 & 0.608 \\ 
$\rho_{\mu \gamma}$ & 0.04         & 0.15          & 0     & 0.08 \\
\hline
\end{tabular}
\end{table}

\begin{figure}[ht]
\includegraphics[width=8.5cm]{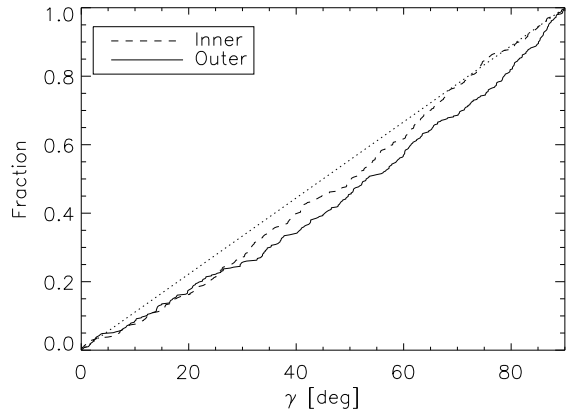}
\caption{Cumulative distributions of angles $\gamma'$ in the inner and
  outer subsystems. The dotted line corresponds to the uniform
  distribution. 
\label{fig:cumgamma} 
}
\end{figure}

\begin{deluxetable*}{l ccc ccc ccc c   }
\tabletypesize{\scriptsize}     
\tablecaption{Eccentricity distributions
\label{tab:e} }  
\tablewidth{0pt}                                   
\tablehead{
\colhead{Case} &                                                                     
\colhead{$f_1$} & 
\colhead{$f_2$} & 
\colhead{$f_3$} & 
\colhead{$f_4$} & 
\colhead{$f_5$} & 
\colhead{$f_6$} & 
\colhead{$f_7$} & 
\colhead{$f_8$} & 
\colhead{$f_9$} & 
\colhead{$f_{10}$} 
}
\startdata
Inner $f_i$       & 0.028& 0.017& 0.019& 0.048& 0.093& 0.141& 0.177& 0.182& 0.160& 0.135 \\
Inner $f_{\rm med}$ & 0.010& 0.020& 0.035& 0.062& 0.097& 0.133& 0.161& 0.174& 0.168& 0.138 \\
Inner $\sigma_f$  & 0.013& 0.010& 0.012& 0.015& 0.016& 0.016& 0.017& 0.017& 0.017& 0.025 \\
Outer $f_i$       & 0.031& 0.018& 0.019& 0.047& 0.087& 0.133& 0.174& 0.183& 0.161& 0.147 \\
Outer $f_{\rm med}$ & 0.010& 0.021& 0.037& 0.063& 0.095& 0.128& 0.156& 0.170& 0.175& 0.144 \\
Outer $\sigma_f$  & 0.013& 0.010& 0.013& 0.015& 0.017& 0.018& 0.019& 0.018& 0.017& 0.024 \\
Outer simulated   & 0.023& 0.070& 0.118& 0.143& 0.142& 0.175& 0.163& 0.114& 0.051& 0.001 \\
\enddata
\end{deluxetable*}

Figure~\ref{fig:cumgamma} shows cumulative distributions of the angles
$\gamma'$. For the  inner subsystems, it is closer  to uniform (dotted
line),  while for  the outer  systems it  is consistently  below.  The
maximum  difference  with  the  uniform  distribution  for  the  outer
systems,  0.110, corresponds  to  the Kolmogorov-Smirnov  significance
level of 0.00012.  The thermal distribution of outer eccentricities is
definitely rejected.

\begin{figure}[ht]
\includegraphics[width=8.5cm]{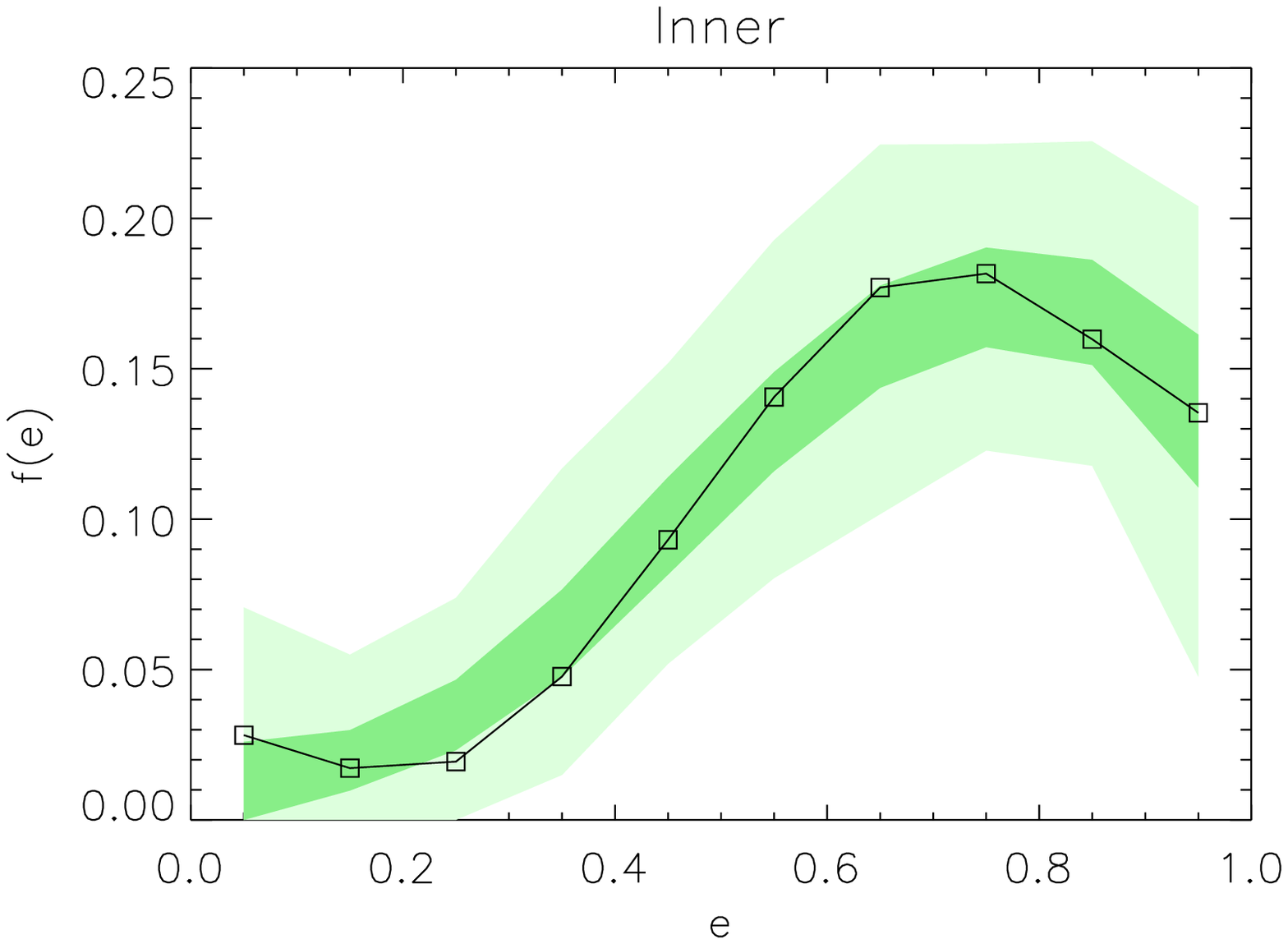} 
\includegraphics[width=8.5cm]{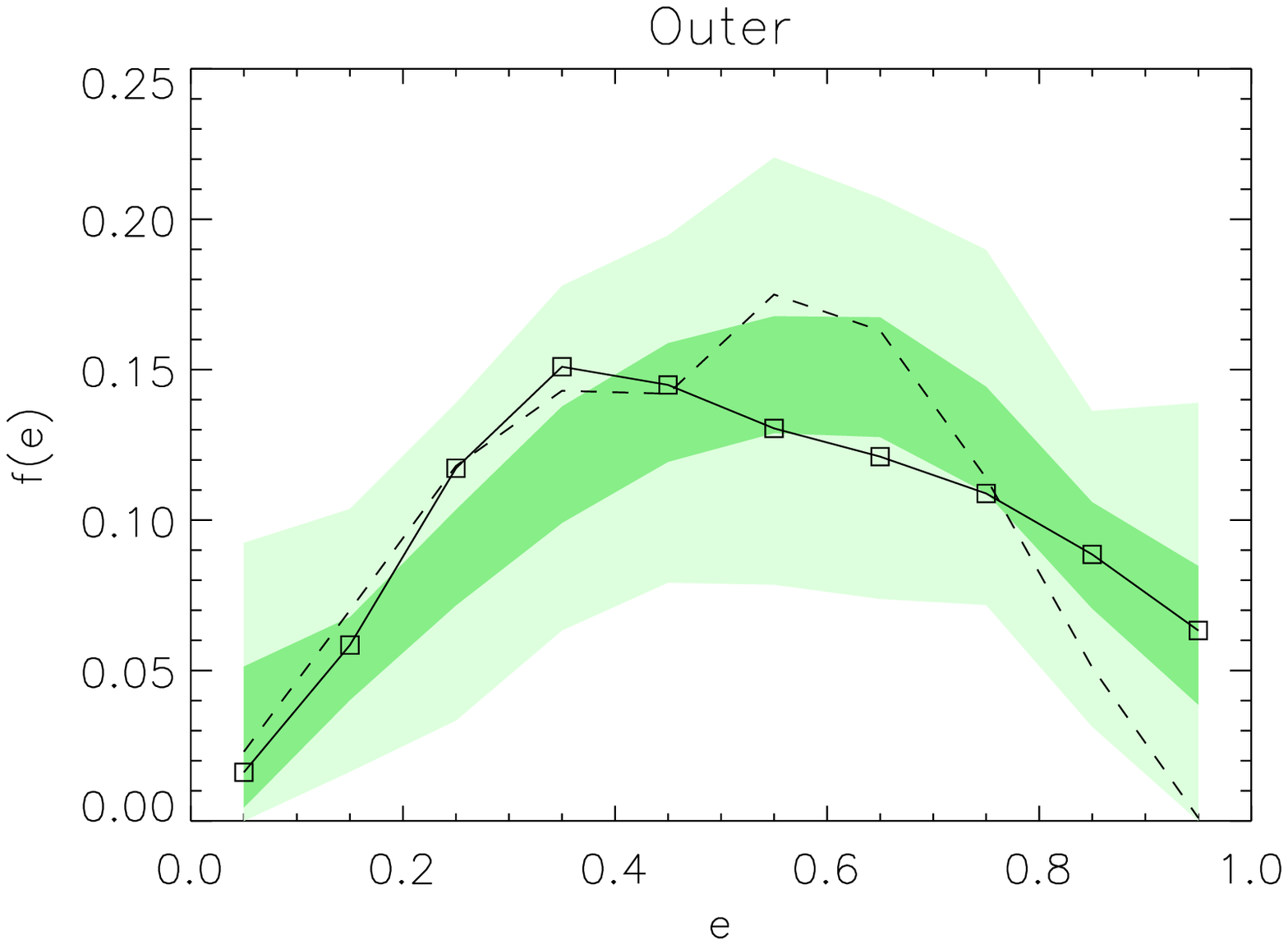} 
\caption{Eccentricity  distributions  in  the inner  (top)  and  outer
  (bottom) subsystems  derived from  the $(\mu',  \gamma')$ statistics
  (squares and full line).  The light and dark green shadings show the
  full and $\pm  \sigma$ range of bootstrap results.   The dashed line
  in the  lower plot  is the eccentricity  distribution expected  from the
  dynamical stability constraint.
\label{fig:fe} 
}
\end{figure}

The eccentricity distributions were  derived from the $(\mu', \gamma')$
arrays by  the algorithm  described in \citet{Tok2020}.   Briefly, the
observed histogram  is modeled  by a  linear combination  of simulated
histograms  (templates) with  coefficients $f_i$.   The templates  are
derived for simulated binaries with  a narrow range of eccentricities,
all  other  orbital  parameters  being random.   I  use  10  templates
corresponding to  0.1 eccentricity  intervals.  The  fitted parameters
$f_i$ are non-negative  and satisfy the condition $\sum_i f_i  = 1$. A
small regularization  parameter $\alpha$  is introduced to  reduce the
noise. After a few trials, I selected $\alpha = 0.01$; $\alpha=0.1$ is
too  large, biasing  the  result toward  uniform distribution,  while
$\alpha  = 0.001$  gives  noisy  results. The  errors  of the  derived
distributions  are  estimated  by  bootstrap:  the  data  are  randomly
re-sampled  1000  times  and  the reconstruction  of  the  eccentricity
distribution is repeated.

The  eccentricity   distributions  of  the  inner   and  outer  orbits
determined by this  method are given in  Table~\ref{tab:e} and plotted
in  Figure~\ref{fig:fe}. The  values $f_i$  (squares connected  by the
full  line) are  obtained directly from the  data. The  full  range of  values
obtained by  bootstrap is shown  by the  light green color,  while the
darker green marks the 68 percentile of the distribution. Half of this
range  gives estimates  of  the  errors $\sigma_f$  and  is listed  in
Table~\ref{tab:e} together with the bootstrap medians $f_{\rm med}$.

The bootstrap analysis  gives the confidence intervals  of the derived
distributions  (which,  however,  depend on  $\alpha$).   The  initial
decreasing part  of the  inner eccentricity  distribution is  likely a
result of random errors because the bootstrap medians demonstrate only
a smooth growth. However, the decline at $e_{\rm in} >0.8$ is a robust
feature. Interestingly, a similar decline was found by \citet{Tok2016}
for wide solar-type binaries (their Figure~7).

The resolution limit of Gaia removes from the sample close inner
subsystems and can affect the resulting distribution of  $(\mu',
\gamma')$ in a systematic way. To address this concern, I simulated a
population of inner binaries by choosing their magnitude differences
$\Delta G$ from the empirical distribution for real inner systems
wider than 2\arcsec, unaffected by the Gaia limit. The semimajor axis
was chosen from a log-uniform distribution between 10 and 330 au, the
distances match the actual distance distribution. Binaries that
happened to be below the Gaia separation-contrast limit were
removed. The bottom panel of Figure~\ref{fig:sep-dg} compares
simulated and real inner subsystems. The template distributions used
in the reconstruction of $f(e)$ for the inner subsystems were derived
from those simulated inner binaries  ($10^4$ per eccentricity
interval). However, the result turned out to be almost identical to
the standard reconstruction that used generic templates without
resolution cutoff. The reconstructed distribution of inner
eccentricities thus appears robust. The mean inner eccentricity is
$e_{\rm in} = 0.66 \pm 0.02$. 

The  distribution of  outer  eccentricities  in Table~\ref{tab:e}  and
Figure~\ref{fig:fe}  is  markedly different  from  the  inner one  and
corresponds  to the  mean  eccentricity  of $e_{\rm  out}  = 0.54  \pm
0.02$. This result is also robust and confirms previous studies: outer
orbits  in hierarchies  are  less eccentric  compared  to binaries  of
similar  separations  \citep{Shatsky2001,Tok2016}.   This  is  natural
because eccentric outer orbits are not allowed by dynamical stability,
especially for  our wide  triples with  moderate separation  ratios. I
show below  that dynamical stability  can indeed explain  the observed
paucity of eccentric outer orbits.

A large number of  outer orbits with a thermal eccentricity
distribution was simulated and unstable systems were removed. The
stability criterion of \citet{Mardling2001} requires  the 
 ratio of semimajor axes $ r = a_{\rm out}/a_{\rm in}$ to be 
larger than the critical ratio $r_{\rm crit}$, 
\begin{equation}
r_{\rm crit} = 2.8 (1 + q_{\rm out})^{1/15} (1 + e_{\rm out})^{0.4} (1 - e_{\rm out})^{-1.2}  .
\label{eq:rcrit}
\end{equation}
For  each system,  a  random  semimajor axis  ratio  was generated  in
agreement with the histogram in Figure~\ref{fig:sepratio}.  The median
outer mass ratio  $q_{\rm out}=0.4$ was adopted.   The resulting outer
eccentricity distribution of the remaining stable triples (dashed line
in  the  lower panel  of  Figure~\ref{fig:fe}  and  the last  line  of
Table~\ref{tab:e})   resembles  the   actual  distribution   of  outer
eccentricities.  The  two curves do  not match exactly because  of 
simplifying assumptions adopted  in the simulations.  I  also used this
simulated $f(e)$ to generate a  simulated $(\mu', \gamma')$ plot.  The
median  values from  this  simulation ($\gamma'  =  56\fdg2$, $\mu'  =
0.61$, $\rho_{\mu  \gamma} =  0.18$) are in  rough agreement  with the
measured values  for outer pairs in  Table~\ref{tab:gamma}.  Thus, the
dynamical stability limit  appears to be the primary  cause of rounded
outer orbits in the resolved triples with moderate separation ratios.

The dynamical stability allows outer  orbits in wide triples to become
more eccentric  at increasing outer separation.   Splitting the sample
into  three groups  by  the  outer separation  (below  1 kau,  $N$=85;
between 1 and  3 kau, $N$=135; and  above 3 kau, $N$=168),  I find the
decreasing   median   $\gamma'$   (58\fdg5,  56\fdg3,   and   49\fdg7,
respectively).  However,  even at  the largest separations,  the outer
eccentricity distribution appears to be softer than the thermal one.

\subsection{Mass ratios}
\label{sec:q}

\begin{figure}[ht]
\includegraphics[width=8.5cm]{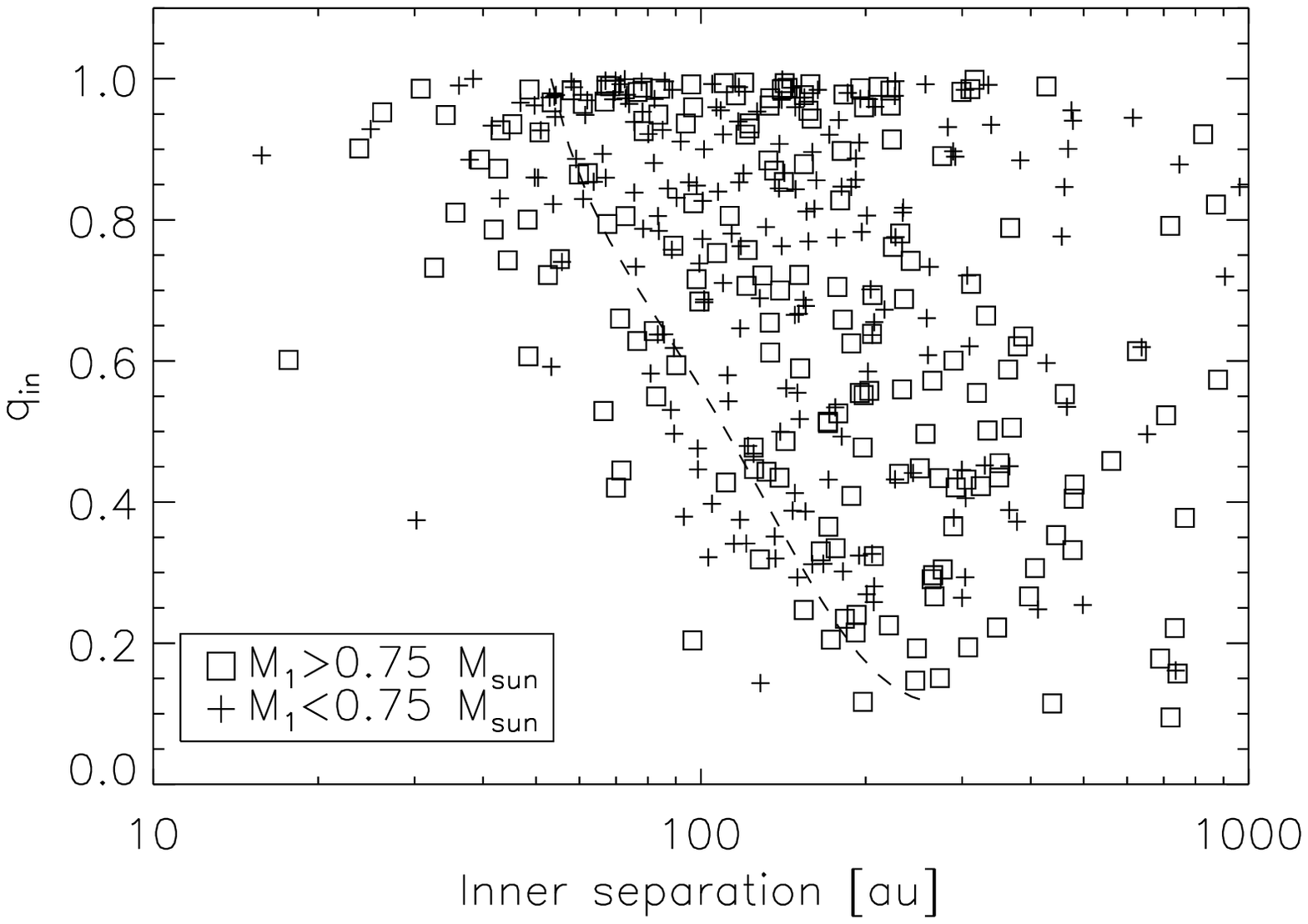} 
\includegraphics[width=8.5cm]{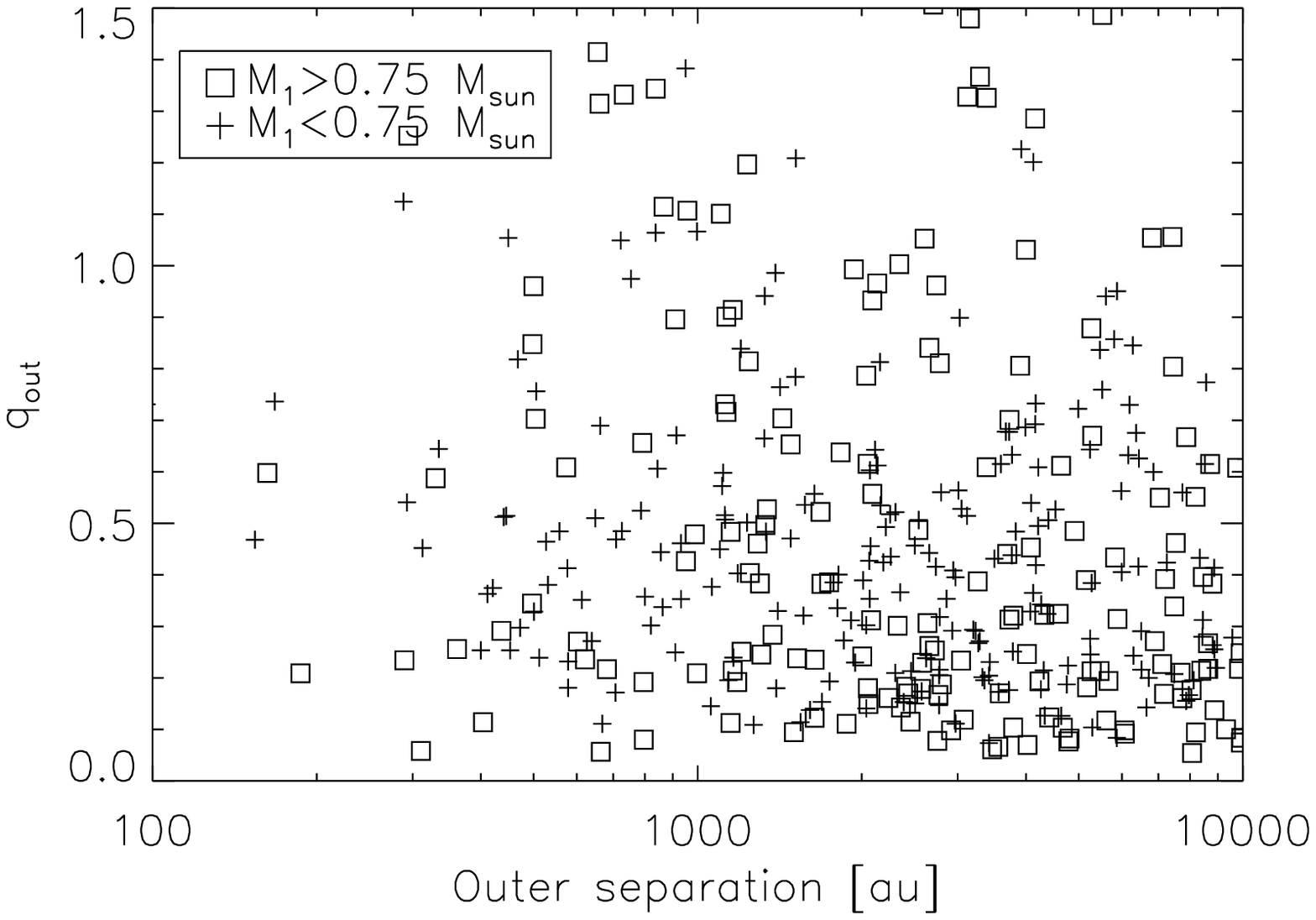} 
\caption{Mass ratio vs. projected separation in the inner (top) and
  outer (bottom) subsystems. The dashed line in the top panel
  illustrates the Gaia detection limit at 100\,pc distance. Squares
  (crosses) denote systems with primaries more (less) massive than
  0.75 \msun.
\label{fig:q-sep} 
}
\end{figure}

The  mass ratios  give additional  insight on  the formation  of these
triples. The inner mass ratio $q_{\rm in} = M_2/M_1$ cannot exceed one
by definition,  while the outer mass  ratio $q_{\rm out} =  M_3/(M_1 +
M_2)$ can be  larger than one.  The masses are  arranged by hierarchy,
not   sorted   in   decreasing   order.    In   the   top   panel   of
Figure~\ref{fig:q-sep},  we  see a  clear  trend  caused by  the  Gaia
detection  limit (high-contrast  pairs  at small  separations are  not
recognized  as distinct  sources).   The dotted  line represents  this
limit (according to eq.~2 in GCNS) at a distance of 100\,pc, converted
into mass  ratio for a  solar-mass primary.  At closer  distance, this
curve is displaced to the left. The envelope of the points follows the
slope of  the detection curve  quite well.   The median values  of the
inner and outer mass ratios are 0.76 and 0.41, respectively.  When the
sample is  split around  the primary  mass of  0.75 \msun,  the median
inner and outer  mass ratios for the low-mass part  are 0.82 and 0.42,
while  for the  high-mass part  they  are slightly  smaller, 0.68  and
0.39. For the 224 triples with inner separation above 2\arcsec, little
affected by the  Gaia detection limit, the median inner  mass ratio is
still high, 0.64.

\begin{figure}[ht]
\includegraphics[width=8.5cm]{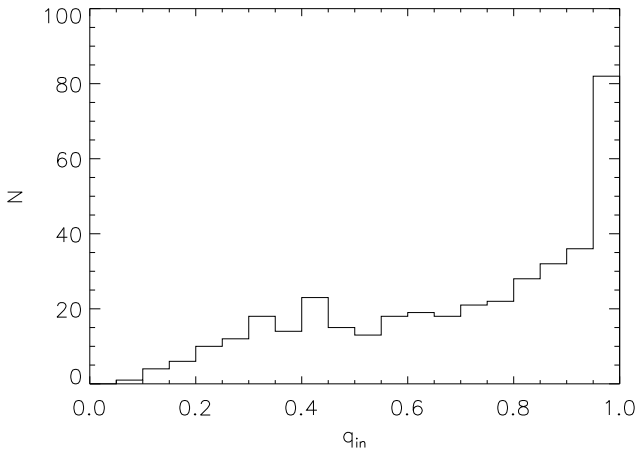} 
\caption{Histogram of inner mass ratios.
\label{fig:qinhist} 
}
\end{figure}

A large number of inner twins with  $q_{\rm in} > 0.95$ is obvious and
it is not  a selection effect, as can be  appreciated in the histogram
of Figure~\ref{fig:qinhist}. The excess of 46 systems in the last bin,
compared  to the  preceding one,  implies the  inner twin  fraction of
$f_{\rm twin}  = 46/392 =  0.12$ according  to the definition  of this
parameter in \citet{Moe2017}.  The  total fraction of inner subsystems
with  $q_{\rm in}  > 0.95$  is  82/392=0.21.  Existence  of wide  twin
binaries  in   the  field   has  been  convincingly   demonstrated  by
\citet{EB2019b}.    Their   Figure~10   gives  the   average   $f_{\rm
  twin}=0.12$ for the  separation range of 50--350 au  and masses from
0.1  to 0.8  \msun,  similar  to the  fraction  found  here for  inner
subsystems.

The outer mass ratios are not  biased by the Gaia detection limit. One
can discern in the bottom panel of Figure~\ref{fig:q-sep} a weak trend
of decreasing outer mass ratios with increasing separation.  Splitting
the sample around 2.7 kau, one  finds the median $q_{\rm out}$ of 0.46
and 0.34 for the closer and wider triples, respectively.

\begin{figure}[ht]
\includegraphics[width=8.5cm]{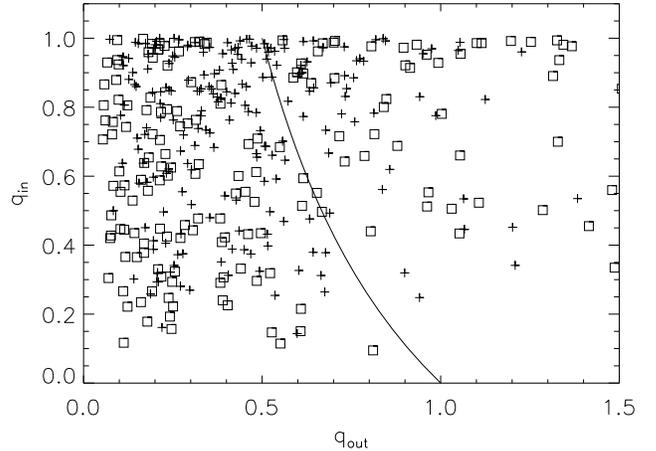} 
\caption{Relation between   inner and outer  mass ratios.  Crosses
  and squares denote triples with  primary masses below and above 0.75
  \msun, respectively. The line corresponds to $M_1 = M_3$.
\label{fig:qlqs} 
}
\end{figure}

Figure~\ref{fig:qlqs} illustrates the relation between inner and outer
mass ratios.  The line $q_{\rm out}  = 1/(1 + q_{\rm in})$ corresponds
to $M_1 =  M_3$ and separates 269 triples where  the most massive star
resides in the inner pair from 123 systems where the most massive star
is the tertiary  component.  So, only 31 per cent  of the triples have
the hierarchy of  A-BC type (a low-mass inner pair  and a more massive
tertiary), while the majority are of AB-C type or AC-B types where the
most massive  star A belongs  to the inner  pair.  Focusing on  the 82
inner twins  with $q_{\rm in}  > 0.95$,  one finds the  median $q_{\rm
  out} = 0.51$, larger than in the full sample. In half of the triples
with  inner twins,  the tertiary  is the  most massive  star.  In  the
remaining 310 triples  without inner twins ($q_{\rm in}  < 0.95$), the
fraction of  massive tertiaries  is 82/310=0.26.   Thus, we  note some
correlation between inner and outer mass ratios.



\section{Origin of Wide Triples}
\label{sec:disc}

Formation of wide hierarchies by turbulent core fragmentation has been
studied by hydrodynamic simulations in several papers. \citet{Lee2019}
show how wide  systems with separations of $\sim$10 kau  form and fall
to the common center of mass  while still accreting.  Those wide pairs
are not  yet binaries because they  have not completed one  full orbit
(the estimated  orbital periods  exceed their age).   Eventually, they
settle into closer orbits. Migration from wide to close separations is
driven   by   two   factors:   mass  accretion   and   dynamical   gas
friction. Unstable groups  of three or more stars  form frequently and
most of  them disintegrate,  ejecting preferentially  low-mass members
(orphans); swaps between components  of hierarchies are also frequent.
Lee et  al.  estimate  that a  0.8 fraction of  all stars  belonged to
systems  at  some  point  in  their history,  even  though  the  final
multiplicity fraction  is only about  0.5. Components of  binaries and
multiples are more  massive than orphans and single  stars (those that
never belonged to any system).

We see that  components of wide triples in the  field are, on average,
more   massive   than   single   stars,  in   agreement   with   these
simulations. This  means that they  have accreted a  substantial mass,
and, consequently, their orbits have  been modified by accretion.  The
stellar system  mass function  (SMF) has a  broad peak  around 0.2-0.3
\msun and a power-law tail at higher masses. The SMF can be modeled by
assuming that low-mass stars representing the SMF peak are produced by
core  collapse,  and  the  high-mass  tail  results  from  competitive
accretion  onto these  low-mass  seeds  \citep{Clark2021}.  From  this
perspective,  our triples  with a  median  system mass  of 1.53  \msun
belong to  the high-mass SMF  tail.  A  substantial fraction of  inner twins
points  to  the   same  conclusion  because  twins   are  produced  by
preferential growth of secondary  components in accreting binaries.  A
binary that formed early and subsequently accreted a major fraction of
its  mass  has  a  high  probability to  become  a  twin.   The  `twin
syndrome', typical for close  binaries \citep{TokMoe2020}, also exists
at  separations  on  the order  of  1  kau  and  beyond, as  shown  by
\citet{EB2019b}.   Formation  of  low-mass   stars  by  core  collapse
produced many twin binaries and multiple systems in the hydrodynamical
simulations   by  \citet{Rohde2021}:   their  Figure~9   qualitatively
resembles    the     histogram    of    inner    mass     ratios    in
Figure~\ref{fig:qinhist},  showing a  preference for  large $q$  and a
prominent twins peak.  \citet{Lomax2015} also predicted preferentially
large mass  ratios. However,  in both  works separations  of simulated
systems are smaller compared to our wide field triples.

The  second  important  conclusion   from  these  simulations  is  the
importance of dynamical interactions  in nascent stellar systems. They
often contain  three or more stars  in non-hierarchical configurations
that decay  by ejecting  orphans.  The   surviving  hierarchies are
dynamically stable,  but not too  far from the stability  limit. Their
orbits are  not aligned, unless  the initial configuration  was highly
flattened and had a  substantial angular momentum \citep{Sterzik2002}.
In the  numerical scattering  experiments where a  binary or  a triple
system interacts with  a passing single or binary  star, the resulting
triples are always near the limit of dynamical stability, their mutual
orbit  orientation is  random,  and the  inner  eccentricities have  a
thermal distribution \citep{Antognini2016}. 

Wide triples  in the field  match the  products of dynamical  decay in
several respects:  their separations are comparable,  the mutual orbit
orientation is almost random, and  the inner eccentricities are large.
However,    the    distribution    of    inner    eccentricities    in
Figure~\ref{fig:fe} is not quite  thermal, flattening and declining at
$e_{\rm in} > 0.7$.  This  can be explained by dissipative interaction
with  gas near  the periastrons  of very  eccentric inner  orbits that
reduces eccentricities to moderate values and contributes to the broad
peak at  $e_{\rm in}  \sim 0.7$.  In the  more compact  visual triples
studied by \citet{Tok2017}, the  inner eccentricities are even smaller
(median 0.5), and there is a  clear tendency to mutual orbit alignment
at outer separations less than 1 kau.  These observations point to the
dynamical importance of gas at separations below $\sim$100 au.  In the
simulations  by \citet{Lomax2015}, triples with aligned orbits and
moderate eccentricities were produced  owing to the energy dissipation
in  gas.  On  the other  hand,  moderate eccentricities  of the  outer
orbits  in  wide  triples  are  a simple  consequence  of  the  N-body
dynamics: only stable triples have survived.

Motions in an  unstable triple lead to ejections of  one star (usually
the smallest) to  large separations. If the system  remains bound, the
ejected     star     returns,     and    the     system     eventually
disintegrates. However,  interactions with gas modify  this picture by
shrinking the  inner orbit,  which may  render the  system dynamically
stable. Interactions of  the ejected star with other  nearby stars can
also stabilize the system by  supplying additional angular momentum to
the  outer  orbit. Most  wide  dynamically  stable triples  formed  by
ejections     (unfolding)     have      eccentric     outer     orbits
\citep{Reipurth2012}.   In contrast,  wide triples  in the  field have
moderate outer  eccentricities, hence the unfolding  mechanism has not
contributed  to their  formation  in any  significant  way.  The  mass
ratios  of  wide  field  triples  do  not  match  the  predictions  of
\citet{Reipurth2015} when we  compare Figure~\ref{fig:qlqs} with their
Figure 14.

Large  outer separations  of the  triples studied  here indicate  that
these systems  have not  experienced disruptive interactions  in dense
clusters   but  rather   formed  in   low-density  environments   like
Taurus-Auriga.  The plot in Figure~\ref{fig:q-sep} suggests that outer
separations  extend  well  beyond  the 10  kau  cutoff  imposed  here.
\citet{Joncour2017} discovered  in Taurus  a population  of ultra-wide
binaries with separations  up to 60 kau, the majority  of them hosting
inner  subsystems   with  separations  below  1   kau.   At  projected
separations  from  1  to  10 kau,  hierarchical  multiples  in  Taurus
outnumber simple  wide binaries.  These authors  determined that stars
belonging  to   hierarchies  are  more  massive   than  single  stars;
components less massive than 0.1  \msun are rare, while typical masses
in  hierarchies are  0.6-0.8  \msun. Joncour  et  al. give  convincing
arguments that wide hierarchies in Taurus are pristine products of the
cascade (hierarchical)  collapse of elongated cores.   These multiples
will  evolve  to  closer  separations  \citep[as  in  the  simulations
  of][]{Lee2019},  some  of  them  will  decay  dynamically,  and  the
surviving systems will possibly resemble the wide triples in the field.

\section{Summary}
\label{sec:sum}

Here are the main results of this study in a condensed form.

\begin{itemize}
\item
A sample of 392 wide low-mass  triples within 100\,pc resolved by Gaia is
selected,   masses  and  internal motions  in  these  systems  are
determined. Most triples contain only three stars, although undetected
inner  subsystems cannot  be fully  excluded. The  sample is  complete
above  the  Gaia separation-contrast  limit;  it  represents a  0.0012
fraction of  the total  field population. The  median primary  mass is
0.71  \msun, and all  components of  these triples  are more  massive than
average field stars.

\item
The  median  inner  and  outer   separations  are  151  and  2570  au,
respectively, and  their median ratio  is 14.75, signaling  that many
wide triples are  just above the dynamical stability  limit. The outer
projected separations are  restricted here to be $<$10  kau, but wider
triples certainly exist in the field.

\item
The inner and outer orbits are aligned almost randomly with the
average mutual inclination of  $83\fdg1 \pm 4\fdg5$. 

\item
The direction  and velocity of  relative motion  is used here to  infer the
eccentricity   distributions.    The   mean  inner   eccentricity   is
0.66$\pm$0.02; its distribution resembles  a thermal one, but flattens
and  declines   at  $e_{\rm  in}   >0.7$.   The  smaller   mean  outer
eccentricity of 0.54$\pm$0.02  can be explained by  the combination of
dynamical stability  and moderate  separation ratios  that effectively
remove eccentric outer orbits from the sample.

\item
The  inner mass  ratios  are  typically large  (median  0.64 at  inner
separation above 2\arcsec), and a  0.21 fraction of inner binaries are
twins with  $q_{\rm in} >  0.95$ (twin  excess of 0.12).   The median
outer mass ratio  $q_{\rm out} = 0.41$ has a  weak trend of decreasing
with outer  separation. Only in 31  per cent of wide  triples the most
massive star is a tertiary component.

\item
Young  wide  hierarchies  in Taurus  studied  by  \citet{Joncour2017},
likely formed by a cascade (hierarchical) collapse of elongated cores,
could be progenitors of wide triples in the field.

\end{itemize}

The present sample of wide triples  represents only a tiny fraction of
all hierarchical systems  within 100\,pc. Assuming that 5  per cent of
stars  more  massive  than  0.3  \msun (the  median  mass)  belong  to
hierarchies,  their total  number within  100\,pc can  be as  large as
8\,000, twenty  times more than this  sample.  Inner pairs in  most of
these hierarchies are not resolved by  Gaia, but their presence can be
often guessed  from the increased  astrometric noise or  variable RVs.
Periods and  mass ratios in  these hypothetical subsystems  remain, so
far, unknown.  Future Gaia data releases will enrich the 100-pc sample
by  adding  close  resolved  binaries and  providing  astrometric  and
spectroscopic orbits  for some  subsystems.  Still, a  tremendous work
remains to be  done to complement Gaia by   high-resolution imaging
and RV monitoring from the ground.  Statistical study of typical, more
compact hierarchical  systems will  help us to  reveal the  mystery of
their formation.

\begin{acknowledgments} 

I thank Bo Reipurth for useful comments on the paper draft.  This work
used the  SIMBAD service operated  by Centre des  Donn\'ees Stellaires
(Strasbourg, France),  bibliographic references from  the Astrophysics
Data System  maintained by SAO/NASA.  This  work has made use  of data
from   the   European   Space   Agency  (ESA)   mission   {\it   Gaia}
(\url{https://www.cosmos.esa.int/gaia}), processed  by the  {\it Gaia}
Data      Processing      and     Analysis      Consortium      (DPAC,
\url{https://www.cosmos.esa.int/web/gaia/dpac/consortium}).    Funding
for the DPAC has been provided by national institutions, in particular
the  institutions   participating  in  the  {\it   Gaia}  Multilateral
Agreement.
\end{acknowledgments}






\begin{thebibliography}{}
\expandafter\ifx\csname natexlab\endcsname\relax\def\natexlab#1{#1}\fi
\providecommand{\url}[1]{\href{#1}{#1}}
\providecommand{\dodoi}[1]{doi:~\href{http://doi.org/#1}{\nolinkurl{#1}}}
\providecommand{\doeprint}[1]{\href{http://ascl.net/#1}{\nolinkurl{http://ascl.net/#1}}}
\providecommand{\doarXiv}[1]{\href{https://arxiv.org/abs/#1}{\nolinkurl{https://arxiv.org/abs/#1}}}

\bibitem[{{Antognini} \& {Thompson}(2016)}]{Antognini2016}
{Antognini}, J. M.~O., \& {Thompson}, T.~A. 2016, \mnras, 456, 4219,
  \dodoi{10.1093/mnras/stv2938}

\bibitem[{{Bate}(2014)}]{Bate2014}
{Bate}, M.~R. 2014, \mnras, 442, 285, \dodoi{10.1093/mnras/stu795}

\bibitem[{{Bressan} {et~al.}(2012){Bressan}, {Marigo}, {Girardi}, {Salasnich},
  {Dal Cero}, {Rubele}, \& {Nanni}}]{PARSEC}
{Bressan}, A., {Marigo}, P., {Girardi}, L., {et~al.} 2012, \mnras, 427, 127,
  \dodoi{10.1111/j.1365-2966.2012.21948.x}

\bibitem[{{Clark} \& {Whitworth}(2021)}]{Clark2021}
{Clark}, P.~C., \& {Whitworth}, A.~P. 2021, \mnras, 500, 1697,
  \dodoi{10.1093/mnras/staa3176}

\bibitem[{{Duch{\^e}ne} \& {Kraus}(2013)}]{DK13}
{Duch{\^e}ne}, G., \& {Kraus}, A. 2013, \araa, 51, 269,
  \dodoi{10.1146/annurev-astro-081710-102602}

\bibitem[{{El-Badry} {et~al.}(2021){El-Badry}, {Rix}, \& {Heintz}}]{EB2021}
{El-Badry}, K., {Rix}, H.-W., \& {Heintz}, T.~M. 2021, \mnras, 506, 2269,
  \dodoi{10.1093/mnras/stab323}

\bibitem[{{El-Badry} {et~al.}(2019){El-Badry}, {Rix}, {Tian}, {Duch{\^e}ne}, \&
  {Moe}}]{EB2019b}
{El-Badry}, K., {Rix}, H.-W., {Tian}, H., {Duch{\^e}ne}, G., \& {Moe}, M. 2019,
  \mnras, 489, 5822, \dodoi{10.1093/mnras/stz2480}

\bibitem[{{Gaia Collaboration} {et~al.}(2016){Gaia Collaboration}, {Brown},
  {Vallenari}, {Prusti}, {de Bruijne}, {Mignard}, \& {et al.}}]{Gaia1}
{Gaia Collaboration}, {Brown}, A.~G.~A., {Vallenari}, A., {et~al.} 2016, \aap,
  595, A2, \dodoi{10.1051/0004-6361/201629512}

\bibitem[{{Gaia Collaboration} {et~al.}(2021{\natexlab{a}}){Gaia
  Collaboration}, {Brown}, {Vallenari}, {Prusti}, {de Bruijne}, {Babusiaux},
  {Biermann}, {Creevey}, {Evans}, {Eyer}, {Hutton}, {Jansen}, {Jordi},
  {Klioner}, {Lammers}, {Lindegren}, {Luri}, {Mignard}, {Panem}, {Pourbaix},
  {Randich}, {Sartoretti}, {Soubiran}, {Walton}, {Arenou}, {Bailer-Jones},
  {Bastian}, {Cropper}, {Drimmel}, {Katz}, {Lattanzi}, {van Leeuwen}, {Bakker},
  {Cacciari}, {Casta{\~n}eda}, {De Angeli}, {Ducourant}, {Fabricius},
  {Fouesneau}, {Fr{\'e}mat}, {Guerra}, {Guerrier}, {Guiraud}, {Jean-Antoine
  Piccolo}, {Masana}, {Messineo}, {Mowlavi}, {Nicolas}, {Nienartowicz},
  {Pailler}, {Panuzzo}, {Riclet}, {Roux}, {Seabroke}, {Sordo}, {Tanga},
  {Th{\'e}venin}, {Gracia-Abril}, {Portell}, {Teyssier}, {Altmann}, {Andrae},
  {Bellas-Velidis}, {Benson}, {Berthier}, {Blomme}, {Brugaletta}, {Burgess},
  {Busso}, {Carry}, {Cellino}, {Cheek}, {Clementini}, {Damerdji}, {Davidson},
  {Delchambre}, {Dell'Oro}, {Fern{\'a}ndez-Hern{\'a}ndez}, {Galluccio},
  {Garc{\'\i}a-Lario}, {Garcia-Reinaldos}, {Gonz{\'a}lez-N{\'u}{\~n}ez},
  {Gosset}, {Haigron}, {Halbwachs}, {Hambly}, {Harrison}, {Hatzidimitriou},
  {Heiter}, {Hern{\'a}ndez}, {Hestroffer}, {Hodgkin}, {Holl}, {Jan{\ss}en},
  {Jevardat de Fombelle}, {Jordan}, {Krone-Martins}, {Lanzafame},
  {L{\"o}ffler}, {Lorca}, {Manteiga}, {Marchal}, {Marrese}, {Moitinho}, {Mora},
  {Muinonen}, {Osborne}, {Pancino}, {Pauwels}, {Petit}, {Recio-Blanco},
  {Richards}, {Riello}, {Rimoldini}, {Robin}, {Roegiers}, {Rybizki}, {Sarro},
  {Siopis}, {Smith}, {Sozzetti}, {Ulla}, {Utrilla}, {van Leeuwen}, {van
  Reeven}, {Abbas}, {Abreu Aramburu}, {Accart}, {Aerts}, {Aguado}, {Ajaj},
  {Altavilla}, {{\'A}lvarez}, {{\'A}lvarez Cid-Fuentes}, {Alves}, {Anderson},
  {Anglada Varela}, {Antoja}, {Audard}, {Baines}, {Baker},
  {Balaguer-N{\'u}{\~n}ez}, {Balbinot}, {Balog}, {Barache}, {Barbato},
  {Barros}, {Barstow}, {Bartolom{\'e}}, {Bassilana}, {Bauchet},
  {Baudesson-Stella}, {Becciani}, {Bellazzini}, {Bernet}, {Bertone}, {Bianchi},
  {Blanco-Cuaresma}, {Boch}, {Bombrun}, {Bossini}, {Bouquillon}, {Bragaglia},
  {Bramante}, {Breedt}, {Bressan}, {Brouillet}, {Bucciarelli}, {Burlacu},
  {Busonero}, {Butkevich}, {Buzzi}, {Caffau}, {Cancelliere}, {C{\'a}novas},
  {Cantat-Gaudin}, {Carballo}, {Carlucci}, {Carnerero}, {Carrasco},
  {Casamiquela}, {Castellani}, {Castro-Ginard}, {Castro Sampol}, {Chaoul},
  {Charlot}, {Chemin}, {Chiavassa}, {Cioni}, {Comoretto}, {Cooper}, {Cornez},
  {Cowell}, {Crifo}, {Crosta}, {Crowley}, {Dafonte}, {Dapergolas}, {David},
  {David}, {de Laverny}, {De Luise}, {De March}, {De Ridder}, {de Souza}, {de
  Teodoro}, {de Torres}, {del Peloso}, {del Pozo}, {Delbo}, {Delgado},
  {Delgado}, {Delisle}, {Di Matteo}, {Diakite}, {Diener}, {Distefano},
  {Dolding}, {Eappachen}, {Edvardsson}, {Enke}, {Esquej}, {Fabre}, {Fabrizio},
  {Faigler}, {Fedorets}, {Fernique}, {Fienga}, {Figueras}, {Fouron},
  {Fragkoudi}, {Fraile}, {Franke}, {Gai}, {Garabato}, {Garcia-Gutierrez},
  {Garc{\'\i}a-Torres}, {Garofalo}, {Gavras}, {Gerlach}, {Geyer}, {Giacobbe},
  {Gilmore}, {Girona}, {Giuffrida}, {Gomel}, {Gomez}, {Gonzalez-Santamaria},
  {Gonz{\'a}lez-Vidal}, {Granvik}, {Guti{\'e}rrez-S{\'a}nchez}, {Guy},
  {Hauser}, {Haywood}, {Helmi}, {Hidalgo}, {Hilger}, {H{\l}adczuk}, {Hobbs},
  {Holland}, {Huckle}, {Jasniewicz}, {Jonker}, {Juaristi Campillo}, {Julbe},
  {Karbevska}, {Kervella}, {Khanna}, {Kochoska}, {Kontizas}, {Kordopatis},
  {Korn}, {Kostrzewa-Rutkowska}, {Kruszy{\'n}ska}, {Lambert}, {Lanza}, {Lasne},
  {Le Campion}, {Le Fustec}, {Lebreton}, {Lebzelter}, {Leccia}, {Leclerc},
  {Lecoeur-Taibi}, {Liao}, {Licata}, {Lindstr{\o}m}, {Lister}, {Livanou},
  {Lobel}, {Madrero Pardo}, {Managau}, {Mann}, {Marchant}, {Marconi}, {Marcos
  Santos}, {Marinoni}, {Marocco}, {Marshall}, {Martin Polo},
  {Mart{\'\i}n-Fleitas}, {Masip}, {Massari}, {Mastrobuono-Battisti}, {Mazeh},
  {McMillan}, {Messina}, {Michalik}, {Millar}, {Mints}, {Molina}, {Molinaro},
  {Moln{\'a}r}, {Montegriffo}, {Mor}, {Morbidelli}, {Morel}, {Morris},
  {Mulone}, {Munoz}, {Muraveva}, {Murphy}, {Musella}, {Noval}, {Ord{\'e}novic},
  {Orr{\`u}}, {Osinde}, {Pagani}, {Pagano}, {Palaversa}, {Palicio}, {Panahi},
  {Pawlak}, {Pe{\~n}alosa Esteller}, {Penttil{\"a}}, {Piersimoni}, {Pineau},
  {Plachy}, {Plum}, {Poggio}, {Poretti}, {Poujoulet}, {Pr{\v{s}}a}, {Pulone},
  {Racero}, {Ragaini}, {Rainer}, {Raiteri}, {Rambaux}, {Ramos}, {Ramos-Lerate},
  {Re Fiorentin}, {Regibo}, {Reyl{\'e}}, {Ripepi}, {Riva}, {Rixon}, {Robichon},
  {Robin}, {Roelens}, {Rohrbasser}, {Romero-G{\'o}mez}, {Rowell}, {Royer},
  {Rybicki}, {Sadowski}, {Sagrist{\`a} Sell{\'e}s}, {Sahlmann}, {Salgado},
  {Salguero}, {Samaras}, {Sanchez Gimenez}, {Sanna}, {Santove{\~n}a},
  {Sarasso}, {Schultheis}, {Sciacca}, {Segol}, {Segovia}, {S{\'e}gransan},
  {Semeux}, {Shahaf}, {Siddiqui}, {Siebert}, {Siltala}, {Slezak}, {Smart},
  {Solano}, {Solitro}, {Souami}, {Souchay}, {Spagna}, {Spoto}, {Steele},
  {Steidelm{\"u}ller}, {Stephenson}, {S{\"u}veges}, {Szabados}, {Szegedi-Elek},
  {Taris}, {Tauran}, {Taylor}, {Teixeira}, {Thuillot}, {Tonello}, {Torra},
  {Torra}, {Turon}, {Unger}, {Vaillant}, {van Dillen}, {Vanel}, {Vecchiato},
  {Viala}, {Vicente}, {Voutsinas}, {Weiler}, {Wevers}, {Wyrzykowski}, {Yoldas},
  {Yvard}, {Zhao}, {Zorec}, {Zucker}, {Zurbach}, \& {Zwitter}}]{Gaia3}
---. 2021{\natexlab{a}}, \aap, 649, A1, \dodoi{10.1051/0004-6361/202039657}

\bibitem[{{Gaia Collaboration} {et~al.}(2021{\natexlab{b}}){Gaia
  Collaboration}, {Smart}, {Sarro}, {Rybizki}, {Reyl{\'e}}, {Robin}, {Hambly},
  {Abbas}, {Barstow}, {de Bruijne}, {Bucciarelli}, {Carrasco}, {Cooper},
  {Hodgkin}, {Masana}, {Michalik}, {Sahlmann}, {Sozzetti}, {Brown},
  {Vallenari}, {Prusti}, {Babusiaux}, {Biermann}, {Creevey}, {Evans}, {Eyer},
  {Hutton}, {Jansen}, {Jordi}, {Klioner}, {Lammers}, {Lindegren}, {Luri},
  {Mignard}, {Panem}, {Pourbaix}, {Randich}, {Sartoretti}, {Soubiran},
  {Walton}, {Arenou}, {Bailer-Jones}, {Bastian}, {Cropper}, {Drimmel}, {Katz},
  {Lattanzi}, {van Leeuwen}, {Bakker}, {Casta{\~n}eda}, {De Angeli},
  {Ducourant}, {Fabricius}, {Fouesneau}, {Fr{\'e}mat}, {Guerra}, {Guerrier},
  {Guiraud}, {Jean-Antoine Piccolo}, {Messineo}, {Mowlavi}, {Nicolas},
  {Nienartowicz}, {Pailler}, {Panuzzo}, {Riclet}, {Roux}, {Seabroke}, {Sordo},
  {Tanga}, {Th{\'e}venin}, {Gracia-Abril}, {Portell}, {Teyssier}, {Altmann},
  {Andrae}, {Bellas-Velidis}, {Benson}, {Berthier}, {Blomme}, {Brugaletta},
  {Burgess}, {Busso}, {Carry}, {Cellino}, {Cheek}, {Clementini}, {Damerdji},
  {Davidson}, {Delchambre}, {Dell'Oro}, {Fern{\'a}ndez-Hern{\'a}ndez},
  {Galluccio}, {Garc{\'\i}a-Lario}, {Garcia-Reinaldos},
  {Gonz{\'a}lez-N{\'u}{\~n}ez}, {Gosset}, {Haigron}, {Halbwachs}, {Harrison},
  {Hatzidimitriou}, {Heiter}, {Hern{\'a}ndez}, {Hestroffer}, {Holl},
  {Jan{\ss}en}, {Jevardat de Fombelle}, {Jordan}, {Krone-Martins}, {Lanzafame},
  {L{\"o}ffler}, {Lorca}, {Manteiga}, {Marchal}, {Marrese}, {Moitinho}, {Mora},
  {Muinonen}, {Osborne}, {Pancino}, {Pauwels}, {Recio-Blanco}, {Richards},
  {Riello}, {Rimoldini}, {Roegiers}, {Siopis}, {Smith}, {Ulla}, {Utrilla}, {van
  Leeuwen}, {van Reeven}, {Abreu Aramburu}, {Accart}, {Aerts}, {Aguado},
  {Ajaj}, {Altavilla}, {{\'A}lvarez}, {{\'A}lvarez Cid-Fuentes}, {Alves},
  {Anderson}, {Anglada Varela}, {Antoja}, {Audard}, {Baines}, {Baker},
  {Balaguer-N{\'u}{\~n}ez}, {Balbinot}, {Balog}, {Barache}, {Barbato},
  {Barros}, {Bartolom{\'e}}, {Bassilana}, {Bauchet}, {Baudesson-Stella},
  {Becciani}, {Bellazzini}, {Bernet}, {Bertone}, {Bianchi}, {Blanco-Cuaresma},
  {Boch}, {Bombrun}, {Bossini}, {Bouquillon}, {Bragaglia}, {Bramante},
  {Breedt}, {Bressan}, {Brouillet}, {Burlacu}, {Busonero}, {Butkevich},
  {Buzzi}, {Caffau}, {Cancelliere}, {C{\'a}novas}, {Cantat-Gaudin}, {Carballo},
  {Carlucci}, {Carnerero}, {Casamiquela}, {Castellani}, {Castro-Ginard},
  {Castro Sampol}, {Chaoul}, {Charlot}, {Chemin}, {Chiavassa}, {Cioni},
  {Comoretto}, {Cornez}, {Cowell}, {Crifo}, {Crosta}, {Crowley}, {Dafonte},
  {Dapergolas}, {David}, {David}, {de Laverny}, {De Luise}, {De March}, {De
  Ridder}, {de Souza}, {de Teodoro}, {de Torres}, {del Peloso}, {del Pozo},
  {Delgado}, {Delgado}, {Delisle}, {Di Matteo}, {Diakite}, {Diener},
  {Distefano}, {Dolding}, {Eappachen}, {Edvardsson}, {Enke}, {Esquej}, {Fabre},
  {Fabrizio}, {Faigler}, {Fedorets}, {Fernique}, {Fienga}, {Figueras},
  {Fouron}, {Fragkoudi}, {Fraile}, {Franke}, {Gai}, {Garabato},
  {Garcia-Gutierrez}, {Garc{\'\i}a-Torres}, {Garofalo}, {Gavras}, {Gerlach},
  {Geyer}, {Giacobbe}, {Gilmore}, {Girona}, {Giuffrida}, {Gomel}, {Gomez},
  {Gonzalez-Santamaria}, {Gonz{\'a}lez-Vidal}, {Granvik},
  {Guti{\'e}rrez-S{\'a}nchez}, {Guy}, {Hauser}, {Haywood}, {Helmi}, {Hidalgo},
  {Hilger}, {H{\l}adczuk}, {Hobbs}, {Holland}, {Huckle}, {Jasniewicz},
  {Jonker}, {Juaristi Campillo}, {Julbe}, {Karbevska}, {Kervella}, {Khanna},
  {Kochoska}, {Kontizas}, {Kordopatis}, {Korn}, {Kostrzewa-Rutkowska},
  {Kruszy{\'n}ska}, {Lambert}, {Lanza}, {Lasne}, {Le Campion}, {Le Fustec},
  {Lebreton}, {Lebzelter}, {Leccia}, {Leclerc}, {Lecoeur-Taibi}, {Liao},
  {Licata}, {Lindstr{\o}m}, {Lister}, {Livanou}, {Lobel}, {Madrero Pardo},
  {Managau}, {Mann}, {Marchant}, {Marconi}, {Marcos Santos}, {Marinoni},
  {Marocco}, {Marshall}, {Martin Polo}, {Mart{\'\i}n-Fleitas}, {Masip},
  {Massari}, {Mastrobuono-Battisti}, {Mazeh}, {McMillan}, {Messina}, {Millar},
  {Mints}, {Molina}, {Molinaro}, {Moln{\'a}r}, {Montegriffo}, {Mor},
  {Morbidelli}, {Morel}, {Morris}, {Mulone}, {Munoz}, {Muraveva}, {Murphy},
  {Musella}, {Noval}, {Ord{\'e}novic}, {Orr{\`u}}, {Osinde}, {Pagani},
  {Pagano}, {Palaversa}, {Palicio}, {Panahi}, {Pawlak}, {Pe{\~n}alosa
  Esteller}, {Penttil{\"a}}, {Piersimoni}, {Pineau}, {Plachy}, {Plum},
  {Poggio}, {Poretti}, {Poujoulet}, {Pr{\v{s}}a}, {Pulone}, {Racero},
  {Ragaini}, {Rainer}, {Raiteri}, {Rambaux}, {Ramos}, {Ramos-Lerate}, {Re
  Fiorentin}, {Regibo}, {Ripepi}, {Riva}, {Rixon}, {Robichon}, {Robin},
  {Roelens}, {Rohrbasser}, {Romero-G{\'o}mez}, {Rowell}, {Royer}, {Rybicki},
  {Sadowski}, {Sagrist{\`a} Sell{\'e}s}, {Salgado}, {Salguero}, {Samaras},
  {Sanchez Gimenez}, {Sanna}, {Santove{\~n}a}, {Sarasso}, {Schultheis},
  {Sciacca}, {Segol}, {Segovia}, {S{\'e}gransan}, {Semeux}, {Shahaf},
  {Siddiqui}, {Siebert}, {Siltala}, {Slezak}, {Solano}, {Solitro}, {Souami},
  {Souchay}, {Spagna}, {Spoto}, {Steele}, {Steidelm{\"u}ller}, {Stephenson},
  {S{\"u}veges}, {Szabados}, {Szegedi-Elek}, {Taris}, {Tauran}, {Taylor},
  {Teixeira}, {Thuillot}, {Tonello}, {Torra}, {Torra}, {Turon}, {Unger},
  {Vaillant}, {van Dillen}, {Vanel}, {Vecchiato}, {Viala}, {Vicente},
  {Voutsinas}, {Weiler}, {Wevers}, {Wyrzykowski}, {Yoldas}, {Yvard}, {Zhao},
  {Zorec}, {Zucker}, {Zurbach}, \& {Zwitter}}]{GCNS}
{Gaia Collaboration}, {Smart}, R.~L., {Sarro}, L.~M., {et~al.}
  2021{\natexlab{b}}, \aap, 649, A6, \dodoi{10.1051/0004-6361/202039498}

\bibitem[{{Joncour} {et~al.}(2017){Joncour}, {Duch{\^e}ne}, \&
  {Moraux}}]{Joncour2017}
{Joncour}, I., {Duch{\^e}ne}, G., \& {Moraux}, E. 2017, \aap, 599, A14,
  \dodoi{10.1051/0004-6361/201629398}

\bibitem[{{Lee} {et~al.}(2019){Lee}, {Offner}, {Kratter}, {Smullen}, \&
  {Li}}]{Lee2019}
{Lee}, A.~T., {Offner}, S. S.~R., {Kratter}, K.~M., {Smullen}, R.~A., \& {Li},
  P.~S. 2019, \apj, 887, 232, \dodoi{10.3847/1538-4357/ab584b}

\bibitem[{{Lomax} {et~al.}(2015){Lomax}, {Whitworth}, {Hubber}, {Stamatellos},
  \& {Walch}}]{Lomax2015}
{Lomax}, O., {Whitworth}, A.~P., {Hubber}, D.~A., {Stamatellos}, D., \&
  {Walch}, S. 2015, \mnras, 447, 1550, \dodoi{10.1093/mnras/stu2530}

\bibitem[{{Mardling} \& {Aarseth}(2001)}]{Mardling2001}
{Mardling}, R.~A., \& {Aarseth}, S.~J. 2001, \mnras, 321, 398,
  \dodoi{10.1046/j.1365-8711.2001.03974.x}

\bibitem[{{Mason} {et~al.}(2001){Mason}, {Wycoff}, {Hartkopf}, {Douglass}, \&
  {Worley}}]{WDS}
{Mason}, B.~D., {Wycoff}, G.~L., {Hartkopf}, W.~I., {Douglass}, G.~G., \&
  {Worley}, C.~E. 2001, \aj, 122, 3466, \dodoi{10.1086/323920}

\bibitem[{{Moe} \& {Di Stefano}(2017)}]{Moe2017}
{Moe}, M., \& {Di Stefano}, R. 2017, \apjs, 230, 15,
  \dodoi{10.3847/1538-4365/aa6fb6}

\bibitem[{{Raghavan} {et~al.}(2010){Raghavan}, {McAlister}, {Henry}, {Latham},
  {Marcy}, {Mason}, {Gies}, {White}, \& {ten Brummelaar}}]{R10}
{Raghavan}, D., {McAlister}, H.~A., {Henry}, T.~J., {et~al.} 2010, \apjs, 190,
  1, \dodoi{10.1088/0067-0049/190/1/1}

\bibitem[{{Reipurth} \& {Mikkola}(2012)}]{Reipurth2012}
{Reipurth}, B., \& {Mikkola}, S. 2012, \nat, 492, 221,
  \dodoi{10.1038/nature11662}

\bibitem[{{Reipurth} \& {Mikkola}(2015)}]{Reipurth2015}
---. 2015, \aj, 149, 145, \dodoi{10.1088/0004-6256/149/4/145}

\bibitem[{{Rohde} {et~al.}(2021){Rohde}, {Walch}, {Clarke}, {Seifried},
  {Whitworth}, \& {Klepitko}}]{Rohde2021}
{Rohde}, P.~F., {Walch}, S., {Clarke}, S.~D., {et~al.} 2021, \mnras, 500, 3594,
  \dodoi{10.1093/mnras/staa2926}

\bibitem[{{Shatsky}(2001)}]{Shatsky2001}
{Shatsky}, N. 2001, \aap, 380, 238, \dodoi{10.1051/0004-6361:20011401}

\bibitem[{{Sterzik} \& {Tokovinin}(2002)}]{Sterzik2002}
{Sterzik}, M.~F., \& {Tokovinin}, A.~A. 2002, \aap, 384, 1030,
  \dodoi{10.1051/0004-6361:20020105}

\bibitem[{{Tokovinin}(2014)}]{FG67a}
{Tokovinin}, A. 2014, \aj, 147, 86, \dodoi{10.1088/0004-6256/147/4/86}

\bibitem[{{Tokovinin}(2017)}]{Tok2017}
---. 2017, \apj, 844, 103, \dodoi{10.3847/1538-4357/aa7746}

\bibitem[{{Tokovinin}(2018)}]{MSC}
---. 2018, \apjs, 235, 6, \dodoi{10.3847/1538-4365/aaa1a5}

\bibitem[{{Tokovinin}(2020)}]{Tok2020}
---. 2020, \mnras, 496, 987, \dodoi{10.1093/mnras/staa1639}

\bibitem[{{Tokovinin} \& {Kiyaeva}(2016)}]{Tok2016}
{Tokovinin}, A., \& {Kiyaeva}, O. 2016, \mnras, 456, 2070,
  \dodoi{10.1093/mnras/stv2825}

\bibitem[{{Tokovinin} \& {Moe}(2020)}]{TokMoe2020}
{Tokovinin}, A., \& {Moe}, M. 2020, \mnras, 491, 5158,
  \dodoi{10.1093/mnras/stz3299}

\bibitem[{{Winters} {et~al.}(2019){Winters}, {Henry}, {Jao}, {Subasavage},
  {Chatelain}, {Slatten}, {Riedel}, {Silverstein}, \& {Payne}}]{Winters2019}
{Winters}, J.~G., {Henry}, T.~J., {Jao}, W.-C., {et~al.} 2019, \aj, 157, 216,
  \dodoi{10.3847/1538-3881/ab05dc}

\end{thebibliography}

\end{document}